% ...

\documentclass[11pt,,multicol]{scrartcl}
\input epsf

\newcounter{figno}
\newcounter{tabno}

\def\crule#1{\multispan{#1}{\hrulefill}} 
\def\tvi#1#2{\vrule height #1pt depth #2pt width 0pt}

\newcommand{\be}{\begin{equation}}
\newcommand{\ee}{\end{equation}}
\newcommand{\bea}{\begin{eqnarray}}
\newcommand{\eea}{\end{eqnarray}}
\newcommand{\nonu}{\nonumber\\}
\newcommand{\bfig}{\begin{figure}}
\newcommand{\efig}{\end{figure}}
\newcommand{\btab}{\begin{table}}
\newcommand{\etab}{\end{table}}
\newcommand{\bcenter}{\begin{center}}
\newcommand{\ecenter}{\end{center}}

\begin{document}
\title{
\vspace*{-1cm}
  \begin{center}
    {\LARGE {\bf Monte Carlo Study of Phase Transitions in the 
        Bond-Diluted 3D 4-State Potts Model}
        }
  \end{center}
}
\author{{\normalsize Christophe Chatelain$^{(a)}$,} 
        {\normalsize Bertrand Berche$^{(a)}$,} \\[-1mm]
        {\normalsize Wolfhard Janke$^{(b)}$} \\[-1mm]
        {\normalsize and Pierre-Emmanuel Berche$^{(c)}$}\\[1mm]
        {\small $^{(a)}$ Laboratoire de Physique des Mat\'eriaux, }
        {\small UMR CNRS 7556,}\\[-1mm]
        {\small Universit\'e Henri Poincar\'e,  Nancy 1,}\\[-1mm]
        {\small F-54506 Vand\oe uvre les Nancy Cedex, France}\\[1mm]
        {\small $^{(b)}$ Institut f\"ur Theoretische Physik,}\\[-1mm]
        {\small Universit\"at Leipzig,}\\[-1mm]
        {\small D-04109 Leipzig, Germany}\\[1mm]
        {\small $^{(c)}$ Groupe de Physique des Mat\'eriaux, }
        {\small UMR CNRS 6634,}\\[-1mm]
        {\small Universit\'e de Rouen,}\\[-1mm]
        {\small  F-76801 Saint Etienne du Rouvray Cedex, France}\\[1mm]
        {\small\tt chatelai@lpm.u-nancy.fr,}\\[-1mm]
        {\small\tt berche@lpm.u-nancy.fr,}\\[-1mm]
        {\small\tt Wolfhard.Janke@itp.uni-leipzig.de,}\\[-1mm]
        {\small\tt Pierre.Berche@univ-rouen.fr} \\[-0.3cm]
}
\maketitle
\vspace{-1cm}
\begin{abstract}
{\small
Large-scale Monte Carlo simulations of the bond-diluted three-dimensional
4-state Potts model are performed. The phase diagram and the physical
properties at the phase transitions are studied using finite-size scaling
techniques. Evidences are given for the existence of a tricritical point
dividing the phase diagram into a regime where the transitions
remain of first order and a second regime where the transitions are softened to
continuous ones by the influence of disorder. In the former regime, the nature
of the transition is essentially clarified through an analysis of the
energy probability distribution. In the latter regime critical exponents are
estimated. Rare and typical events are identified and their role 
is qualitatively discussed in both regimes. 
\\[0.2cm]
{\bf PACS:}\\
{\em 05.40.+j Fluctuation phenomena, random processes, and Brownian motion;\\
64.60.Fr Equilibrium properties near critical points, critical exponents;\\
75.10.Hk Classical spin models  
}}
\end{abstract}

\section{Introduction\label{sec1}}
The influence of disorder is of great interest in physics, since pure systems
are rare in nature. It has been known for more than 
thirty years that the universality
class associated with a continuous phase transition can be changed by the
presence of quenched impurities~\cite{Khmelnitskii74}. According to the Harris
criterion~\cite{Harris74}, uncorrelated randomness coupled to the energy density
can only affect the critical behaviour of a system if the critical exponent $\alpha$
describing the divergence of the specific heat in the pure system is
positive. This has been established in the case of the $q$-state Potts model in
dimension $D=2$ for example. For $2<q\le 4$, the pure system undergoes a
continuous transition with a positive critical exponent $\alpha$.
As predicted by the Harris criterion, new universality classes have been
observed both perturbatively and numerically~\cite{Potts2D} (for a review, see
Ref.~\cite{BercheChatelain03}). The special case $q=2$, the Ising model, is
particularly interesting since in the pure system, the specific heat displays
a logarithmic divergence ($\alpha=0$) making the Harris criterion inconclusive~\cite{DotsDots83}.
Based on perturbative and numerical studies, it is now generally believed that
the critical behaviour remains unchanged apart from logarithmic corrections when
introducing randomness in the system~\cite{Ising2D}.
In three dimensions (3D), the disordered Ising model was subject of really 
extensive studies (see, e.g., Ref.~\cite{FolkEtAl03} for an exhaustive list of
references).

Less attention has been paid to first-order phase transitions. It is known that
randomness coupled to the energy density softens any temperature-driven 
first-order phase transition~\cite{ImryWortis79}. Moreover, it has been rigorously
proved~\cite{AizenmanWehr89HuiBerker89} that in dimension $D\le 2$ an infinitesimal
amount of disorder is sufficient to {\em turn any first-order transition into 
a continuous one}. The first observation of such a change of the order of the
transition was made in the 2D 8-state Potts model~\cite{ChenFerrenbergLandau92} 
where a new universality class was identified~\cite{cardy_jacobsen,Pottsq8}. For higher
dimensions, the first-order nature of the transition may persist up to a finite amount
of disorder. A tricritical point at finite disorder between two regimes of respectively
first-order and continuous transitions is expected \cite{cardy_jacobsen,cardy}.
The existence of such a tricritical point for the site-diluted 3D 3-state Potts model
could only be suspected by simulations because the pure model already undergoes
a very weak first-order phase transition~\cite{BallesterosEtAl00}.
On the other hand, the first-order phase transition of the pure 5-state
Potts model is very strong and would hence make it rather difficult to study the
role of disorder. As a consequence, we have turned our attention
to the 3D 4-state Potts model and have shown that there exists a second-order transition
regime for this model~\cite{ChatelainEtAl01}. Our choice of bond dilution is motivated
by the fact that for this model only high-temperature expansions results are available up to now
which to our knowledge cannot be done for site-dilution or are at least more
difficult~\cite{HellmundJanke02}.
 
In Sect.~\ref{sec2} we define the model and the observables, and remind
the reader of how these quantities behave at first- and second-order phase transitions. 
Section~\ref{sec3} is devoted to the numerical procedure, first the description of and
then the comparison between the algorithms which are used at low and
high impurity concentrations, followed by a first discussion of the 
qualitative properties of the disorder average.
A short characterisation of the nature of the phase transition -- 
at a qualitative level -- is
reported in Sect.~\ref{sec4}. The motivation of this section is to first
convince ourselves that the transition does indeed undergo a 
qualitative change when the strength of disorder is varied.
Then,  we describe how the phase diagram is obtained  and concentrate on
the first-order regime in Sect.~\ref{sec5}. In Sect.~\ref{sec6}, 
we discuss the critical behaviour in the second-order induced regime. Finally,
the main features of the paper are summarised in Sect.~\ref{sec7}.

\section{Model and observables\label{sec2}}
We study the disordered 4-state Potts model on a  cubic lattice $\Lambda$. 
The model is defined by the Hamiltonian
\begin{equation}
  {\cal H}[\sigma,J]=-\sum_{(i,j)} J_{ij}\delta_{\sigma_i,\sigma_j},
  \label{eq1}
\end{equation}
where the spins $\sigma_i$, located on the vertices $i$ of the lattice 
$\Lambda$, are allowed to take one of the $q=4$ values $\sigma_i=1,\dots,q$. 
The boundaries are chosen periodic in the three space directions.
The notation ${\cal H}[\sigma,J]$ specifies that the Hamiltonian is defined
for  any  configuration of spins and of couplings. The sum runs over the
couples of nearest-neighbouring sites and the exchange couplings
$J_{ij}$ are independent quenched, random variables, distributed according to
the normalised binary distribution ($J > 0$)
\begin{equation}
  P[J_{ij}]=\prod_{(i,j)}[p\delta(J_{ij}-J)+(1-p)\delta(J_{ij})].
  \label{eq2}
\end{equation}
The pure system (at $p=1$) undergoes a strong first-order phase transition with a
correlation length $\xi\sim 3$ lattice units at the inverse transition
temperature $\beta_cJ=0.628\,63(2)$~\cite{JankeKappler96}
 (we keep the conventional notation
$\beta=(k_BT)^{-1}$, since in the context
there is no risk of confusion with the critical exponent of the 
magnetisation). 
As far as we know, no more information has
been made available on this model. We do not expect any phase transition for
bond concentration $p$ smaller than the percolation threshold $p_c =
0.248\ \!812\ \!6(5)$~\cite{LorenzZiff97} 
since the absence of a percolating cluster makes
the appearance of long-range order impossible.

In the following, we are thus dealing with quenched dilution. The 
averaging prescription
is such that
the physical quantities of interest  in the diluted system (say an 
observable $Q$)
are obtained after {\em averaging first a given sample $[J]$ over the
Boltzmann distribution, $\langle Q_{[J]}\rangle_\beta$,  and then over 
the  random distribution of the couplings} denoted by
$\overline{\langle Q_{[J]}\rangle_\beta}$, 
since there is no thermal relaxation of
the degrees of freedom associated to quenched disorder:

\begin{itemize}
\item[$\triangleright$]
The thermodynamic average of an observable $Q$ at inverse temperature
$\beta$ and for a given disorder realization $[j]$ is denoted 
\bea
\langle Q_{[J]}\rangle_\beta&=&(Z_{[J]}(\beta))^{-1}\int {\cal D}
[\sigma]Q_{[\sigma,J]}{\rm e}^{-\beta{\cal H}[\sigma,J]}\nonu
&\approx&\frac{1}{N_{\rm MCS}}\sum_{{\rm MCS}}Q_{[J]}(\beta),
\label{eq2bis}
\eea
where $N_{\rm MCS}$ is the number of Monte Carlo iterations (Monte Carlo
steps) during the ``production'' part after the system has been thermalised.
Here we use the following notation: $Q_{[\sigma,J]}$ is the value of $Q$
for a given spin configuration $[\sigma]$ and a given disorder realization $[J]$,
$Q_{[J]}(\beta)$ is a value obtained by Monte Carlo simulation at inverse
temperature $\beta$. Time to time, we will have to specify a particular disorder 
realization, say $\# n$, and the value of the observable $Q$ for this very
sample will be denoted as  $Q_{\#n}(\beta)$.

\item[$\triangleright$]
The average over randomness is then performed,
\bea
  \overline{\langle Q_{[J]}\rangle_\beta}&=&\int {\cal D}[J] \langle 
        Q_{[J]}\rangle_\beta P[J] \cr
        &\approx&\frac{1}{N\{J\}}\sum_{[J]}\langle Q_{[J]}\rangle_\beta  \cr
        &=&\int {d}\ \!Q\ \!Q P_\beta(Q),
\label{eq2ter}
\eea 
where $N\{J\}$ is the number of independent samples. The probability
$P_\beta(Q)$ is determined empirically from the discrete set of values
of $\langle Q_{[J]}\rangle_\beta$.
This disorder average is  simply denoted as $\overline Q(\beta)$ for short,
i.e. $\overline Q(\beta)\equiv\overline{\langle Q_{[J]}\rangle_\beta}$.
\end{itemize}

For a specific disorder realization $[J]$, the magnetisation per
spin $m_{[\sigma,J]} = L^{-D} M_{[\sigma,J]}$ of the
spin configuration $[\sigma]$ is defined from the fraction 
of spins, $\rho_{[\sigma,J]}$, that are in
the majority orientation,
\begin{eqnarray}
  \rho_{[\sigma,J]}&=&\max_{\sigma_0}\left[L^{-D}\sum_{i\in\Lambda}
\delta_{\sigma_i,\sigma_0}\right],\cr
  m_{[\sigma,J]}&=&\frac{q\rho_{[\sigma,J]}-1}{q-1}.
  \label{eq3}
\end{eqnarray}
The order parameter of the diluted system is thus denoted 
$\overline m(\beta)=\overline{\langle m_{[J]}\rangle_\beta}$. 
Thermal and disorder moments
$\overline{\langle m_{[J]}^n\rangle_\beta}$ and 
$\overline{\langle m_{[J]}\rangle_\beta^n}$, respectively, are also
quantities of interest. The magnetic susceptibility 
$\chi_{[J]}(\beta)$ and the specific heat $C_{[J]}(\beta)$ of a sample 
are defined using the fluctuation-dissipation theorem, i.e.
\bea
  \chi_{[J]}(\beta)&=&\beta L^D\left[{\langle m_{[J]}^2\rangle_\beta}
    -{\langle m_{[J]}\rangle_\beta^2}\right],  \label{eq4Chi}\\[1mm]
    C_{[J]}(\beta)/k_B&=& \beta^2 L^D\left[{\langle e_{[J]}^2\rangle_\beta}
     -{\langle e_{[J]}\rangle^2_\beta}\right],
  \label{eq4C}
\eea
where 
\begin{equation}
 	e_{[\sigma,J]} = L^{-D} E_{[\sigma,J]}
 	=L^{-D}\sum_{(i,j)} J_{ij}
          \delta_{\sigma_{i},\sigma_{j}}.
          \label{eqE}
\end{equation}
is the negative energy density since $E_{[\sigma,J]}=-{\cal H}[\sigma,J]$.
Binder cumulants~\cite{Binder81} take their usual definition, for example
\begin{equation}
  U_{m_{[J]}}(\beta)=
        1-{{{\langle m_{[J]}^4\rangle_\beta}\over
        3{\langle m_{[J]}^2\rangle_\beta}^2}}.
  \label{eq5}
\end{equation}
Derivatives with respect to the exchange coupling are computed
through
\begin{equation}
   L^{-D}{d\over d\beta}\ln \langle m_{[J]}^n\rangle_\beta
  	= {{\langle m_{[J]}^ne_{[J]}\rangle_\beta
	\over \langle m_{[J]}^n\rangle_\beta}
      	-\langle e_{[J]}\rangle_\beta}.
  \label{eq6}
\end{equation}

All these quantities are then averaged over disorder, yielding
$\overline\chi(\beta)$, $\overline C(\beta)$, $\overline U_m(\beta)$,
and $\overline{\partial_\beta\ln\langle m_{[J]}^n\rangle_\beta}$.

At a second-order transition, these quantities are expected to exhibit 
singularities described in terms of power laws from the
deviation to the critical point. These power laws define the critical 
exponents.
In the following, the properties will be investigated using 
finite-size scaling analyses, i.e., according to the following size
dependence at the critical temperature,
\begin{eqnarray}
\overline m({\beta_c},L^{-1})&\sim&B_c L^{-\beta/\nu},\label{eq7bisa}\\[1mm]
\overline\chi({\beta_c},L^{-1})&\sim&\Gamma_c L^{\gamma/\nu},
        \label{eq7bisb}\\[1mm]
\overline C({\beta_c},L^{-1})&\sim&A_c L^{\alpha/\nu},\label{eq7bisc}\\[1mm]
L^{-D}\left.\overline {d\ln \langle m_{[J]}^n\rangle_\beta\over d\beta}
\right|_{\beta_c}&\sim&N_{n,c} L^{1/\nu}.\label{eq7bisd}
\end{eqnarray}

At a first-order transition, the order parameter has a discontinuity at 
the transition
temperature, suggesting that $\beta/\nu$ formally becomes zero. Heuristic (and
for pure $q$-state Potts models with sufficiently large $q$ even
rigorous) arguments
also suggest that $\gamma/\nu$, $\alpha/\nu$, and $1/\nu$ should then 
coincide with the space 
dimension $D$~\cite{BorgsImbrie89BorgsKotecky90Cabrera90}, restoring
the ordinary extensivity of the system in Eqs.~(\ref{eq7bisb}) -- 
(\ref{eq7bisd}).

When the transition temperature is not known exactly, the problem of the
value of the inverse critical temperature
$\beta_c$ in the  expressions above can be a source 
of further difficulties.
Usually, one follows a flow of finite-size estimates given by the location of 
the maximum of a diverging quantity (for example the susceptibility, 
$\overline\chi_{\rm max}(L^{-1})\equiv\ \!{\rm max}_\beta[\overline 
\chi(\beta,L^{-1})]$). 
From the
scaling assumption, supposed to apply at the random fixed point, 
\be
\overline \chi(\beta,L^{-1})=L^{\gamma/\nu}f_\chi(
L^{1/\nu} t),
\ee 
with $t=|\beta_c-\beta|$, the inverse temperature $\beta_{\rm max}$ where 
the maximum of $\bar\chi$ occurs,
\be
\overline\chi(\beta_{\rm max},L^{-1})=\overline\chi_{\rm max}(L^{-1}),
\label{EqDfnBetaMax}
\ee 
scales according to 
\be\beta_{\rm max}\sim \beta_c + aL^{-1/\nu}.\ee
Notice that the scaling
function $f_\chi(x)$ takes its maximum value
$f_\chi(a)=\overline\chi_{\rm max}L^{-\gamma/\nu}$ at $x=a$.

At that very temperature
where the finite-size susceptibility has its maximum, we then have
similar power law expressions,
\begin{eqnarray}
\overline  m({\beta_{\rm max}},L^{-1})
&\sim&f_m(a) L^{-\beta/\nu},\label{eq7tera}\\[1mm]
\overline\chi({\beta_{\rm max}},L^{-1})&\sim&f_\chi(a) L^{\gamma/\nu},
        \label{eq7terb}\\[1mm]
\overline C({\beta_{\rm max}},L^{-1})&\sim&f_C(a) L^{\alpha/\nu},
        \label{eq7terc}\\[1mm]
 L^{-D}\left.\overline {d\ln \langle m_{[J]}^n\rangle_\beta\over d\beta}
\right|_{\beta_{\rm max}}&\sim&f_{m,n}(a) L^{1/\nu}.\label{eq7terd}
\end{eqnarray}
These equations are similar to Eqs.~(\ref{eq7bisa}) -- (\ref{eq7bisd}) 
where the amplitudes
take the values $B_c=f_m(0)$,
$\Gamma_c=f_\chi(0)$, $A_c=f_C(0)$, and $N_c=f_{m,n}(0)$.

From this discussion, we are led to give a more precise definition of
$\overline\chi_{\rm max}(L^{-1})$. One reasonable alternative definition of this 
quantity could be the disorder average of the {\em individual maxima\/}
corresponding to the different samples. Each of them has its own susceptibility curve,
$\chi_{[J]}(\beta,L^{-1})$ which displays a maximum $\chi_{[J],{\rm max}}(L^{-1})$
at a given value of the inverse temperature $\beta_{[J]}^{\rm max}(L^{-1})$.
These values $\chi_{[J],{\rm max}}(L^{-1})$
may then be averaged, but this is in general different from the definition
that we gave for the average over randomness in Eq.~(\ref{eq2ter}). 
Here we keep as a physical quantity the
expectation value $\overline\chi(\beta,L^{-1})$ which is then plotted
against $\beta$, and $\beta_{\rm max}(L^{-1})$ in Eq.~(\ref{EqDfnBetaMax}) 
is the temperature where the disorder
averaged susceptibility displays its maximum which is thus identified
with $\overline\chi_{\rm max}(L^{-1})$.
In the following, this is the physical content that we understand when discussing
$\overline\chi_{\rm max}(L^{-1})$.

\section{Numerical procedures\label{sec3}}
We conducted a long-term and extensive study of the bond-diluted 3D 4-state Potts model,
and it is the purpose of this paper to report results for moderately large system sizes
in the first-order regime, and an extended analysis based on really large-scale computations%
in the second-order regime. 
Cross-over effects between different regimes are also discussed.
The simulations were performed on the significant scale of
several years. A strict organisation was thus required, and we proceeded
as follows: as an output of the runs, all the data were stored in a binary 
format. For each sample (with a given disorder realization and lattice size)
and each simulated temperature, the time series of the energy and magnetisation
were stored. A code was written in order to extract from all
the available files the histogram reweightings of thermodynamic quantities
of interest, entering as an 
input the chosen dilution, lattice size, temperature, 
\dots. It is also possible to adjust the number of thermalisation iterations,
the length of the production runs 
where the thermodynamic averages are performed,
the number of samples for the disorder average, or to pick a specific 
disorder realization, and so on. In some sense, {\em the time series 
correspond to the simulation of the system, and we can then measure
physical quantities on it}, and virtually produce as many results as we
want. Of course, this is {\it not} what we intend to do in the following, 
we rather shall try to concentrate only on the most important results.

\subsection{Choice of update algorithms}
We studied this system by really large-scale Monte Carlo simulations. 
A preliminary study, needed in order to schedule such large-scale Monte Carlo 
simulations, showed that the transitions at small and high 
concentrations of non-vanishing bonds $p$ were (as expected)
qualitatively different:
\begin{itemize} 
\item[$\triangleright\ $]
Close to the pure system, $p\simeq 1$, 
the susceptibility peaks develop as the size increases to become quite 
sharp (see Fig.~\ref{Fig2chi}), in agreement with
what is expected at a first-order phase
transition~\cite{MeyerOrtmannsReisz98WJ03}. 
\item[$\triangleright\ $] 
At larger dilutions (small values of $p$)
on the other hand, the peaks are softened and are compatible, at least at 
first sight, with a second-order phase transition (see Fig.~\ref{Fig2chi}).
\end{itemize}

\bfig[tb]
        \begin{center}
        \epsfxsize=9.5cm
        \mbox{\epsfbox{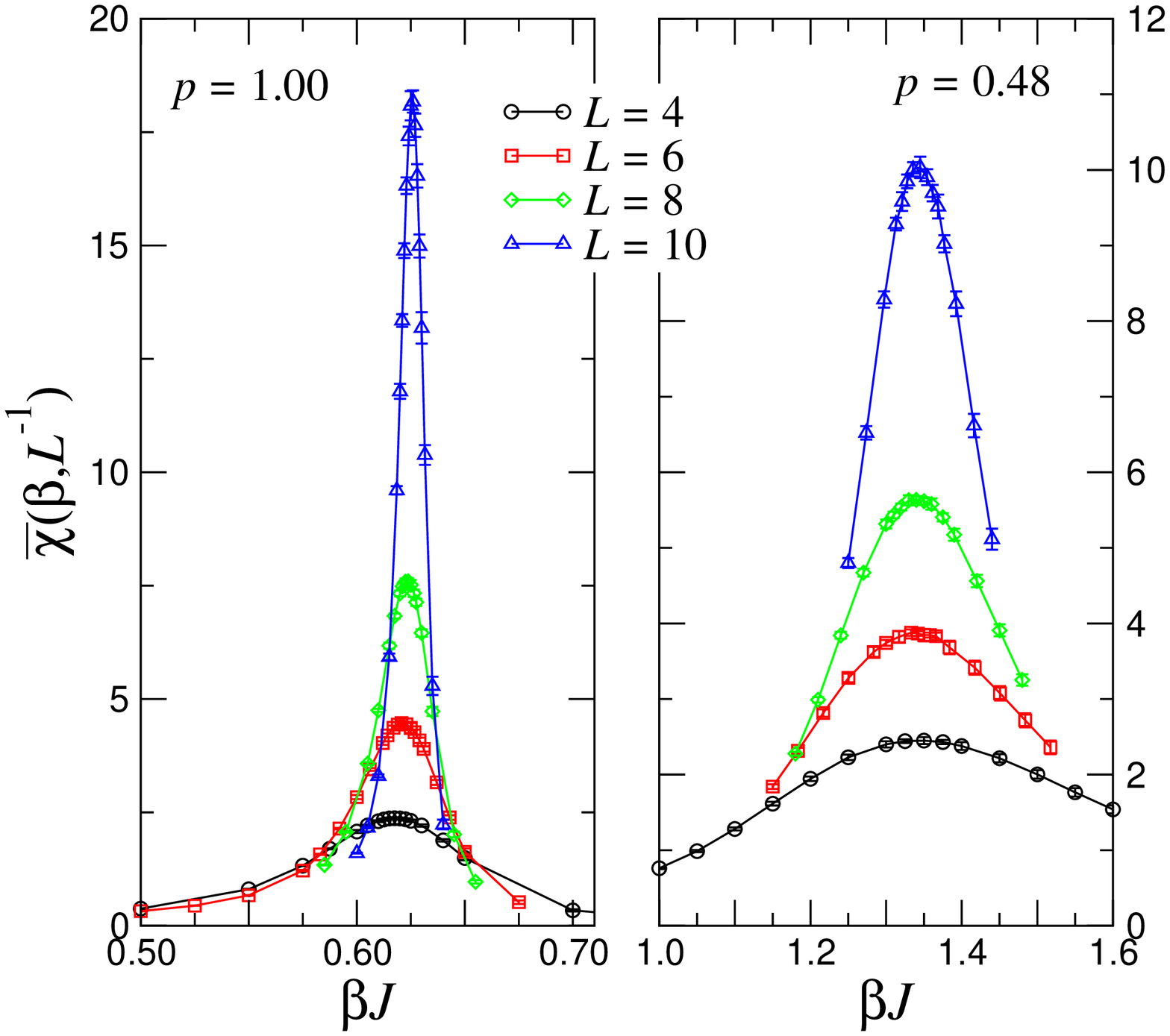}}
        \end{center}\vskip 0cm
        \caption{\small Evolution of the susceptibility as the size of the 
        system increases (up to $L=10$) in the two
        different regimes: pure system $p=1.0$ on the left plot and 
        high dilution $p=0.48$ on the right plot.}
        \label{Fig2chi}
\efig

As will be demonstrated below, the tricritical dilution dividing these 
two regimes is roughly located at $p_{\rm TCP} \approx 0.68 - 0.84$.
In the regime of randomness-induced continuous transitions (or weak first-order 
transitions, that is at low non-zero bond concentration $p$), the Swendsen-Wang
cluster algorithm~\cite{SwendsenWang87} was preferred in order to reduce the
critical slowing-down.
As already pointed out by Ballesteros
{\sl et al.}~\cite{BallesterosEtAl98, BallesterosEtAl00}, a typical spin
configuration at low bond concentration is composed of disconnected clusters
for most of the disorder realisations. It is thus safer to use the Swendsen-Wang
algorithm, for which the whole lattice is swept at each Monte Carlo iteration,
instead of a single-cluster Wolff update procedure.
In the strong first-order regime (high bond concentration $p$), the multi-bondic
algorithm~\cite{JankeKappler95}, a multi-canonical version of the Swendsen-Wang
algorithm, was chosen in order to enhance tunnellings between the phases in
coexistence at the transition temperature. The Swendsen-Wang algorithm,
being less time-consuming, was nevertheless preferred even in this regime of long thermal
relaxation as long as at least ten tunnelling events
between the ordered and disordered phases could be 
observed. As the first-order regime is approached, more and more sweeps
are needed to fulfil this condition. We had to use up to
\begin{description}
\item{--} $200\ \!000$  Monte Carlo steps (MCS) at $p=0.76$ for $L=16$ for example, 
\end{description}
while in the second-order regime for much larger systems, we needed 
\begin{description}
\item{--} $100\ \!000$~MCS
at $p=0.68$ for $L=50$,
\item{--} $30\ \!000$ MCS at $p=0.56$ for $L=96$, and
\item{--} $15\ \!000$ MCS at $p=0.44$  for $L=128$.
\end{description}

\begin{figure*}
        \begin{center}
        \epsfxsize=9.5cm
        \mbox{\epsfbox{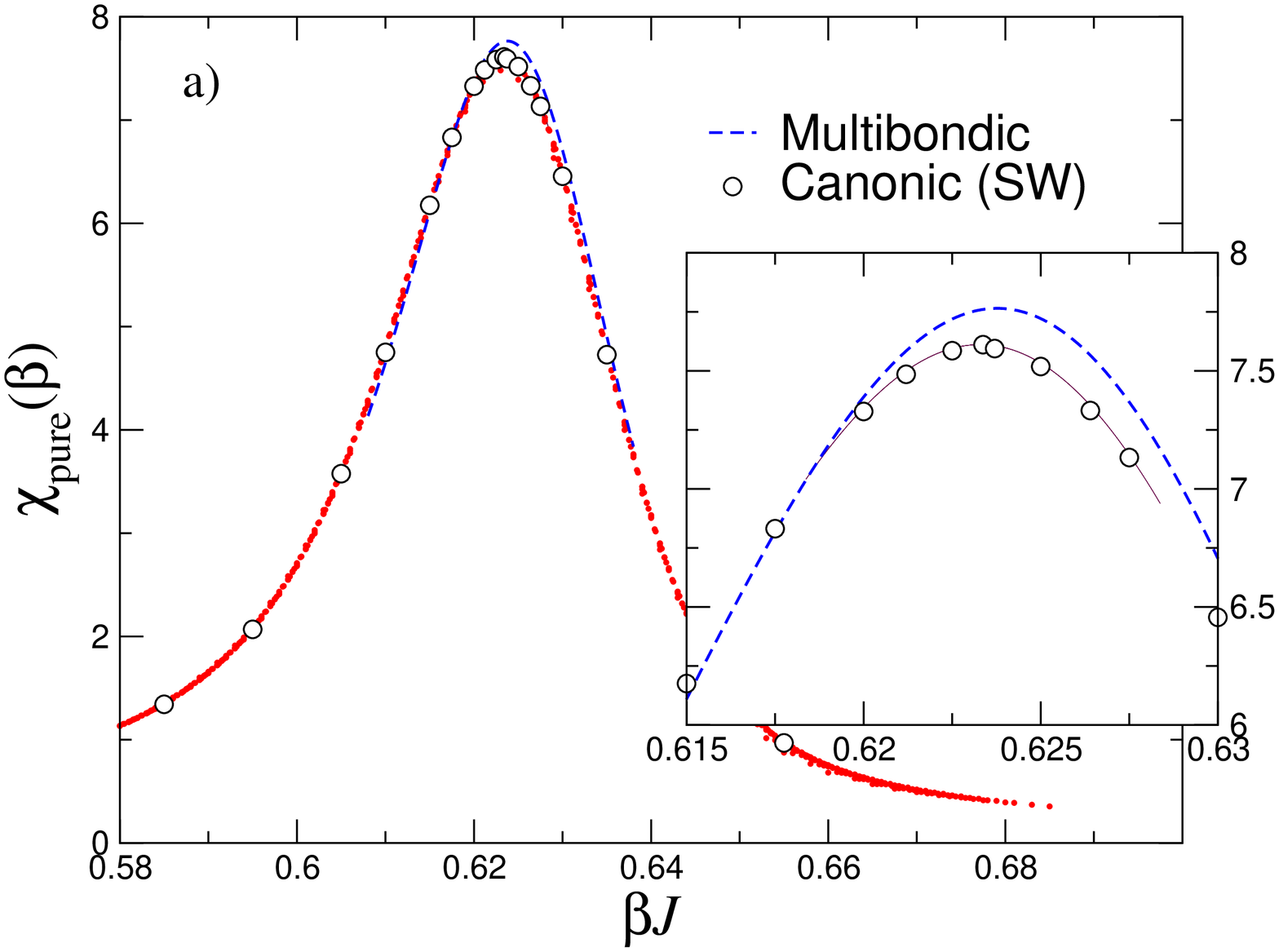}}
        \vskip 0cm
        \epsfxsize=9.5cm
        \mbox{\epsfbox{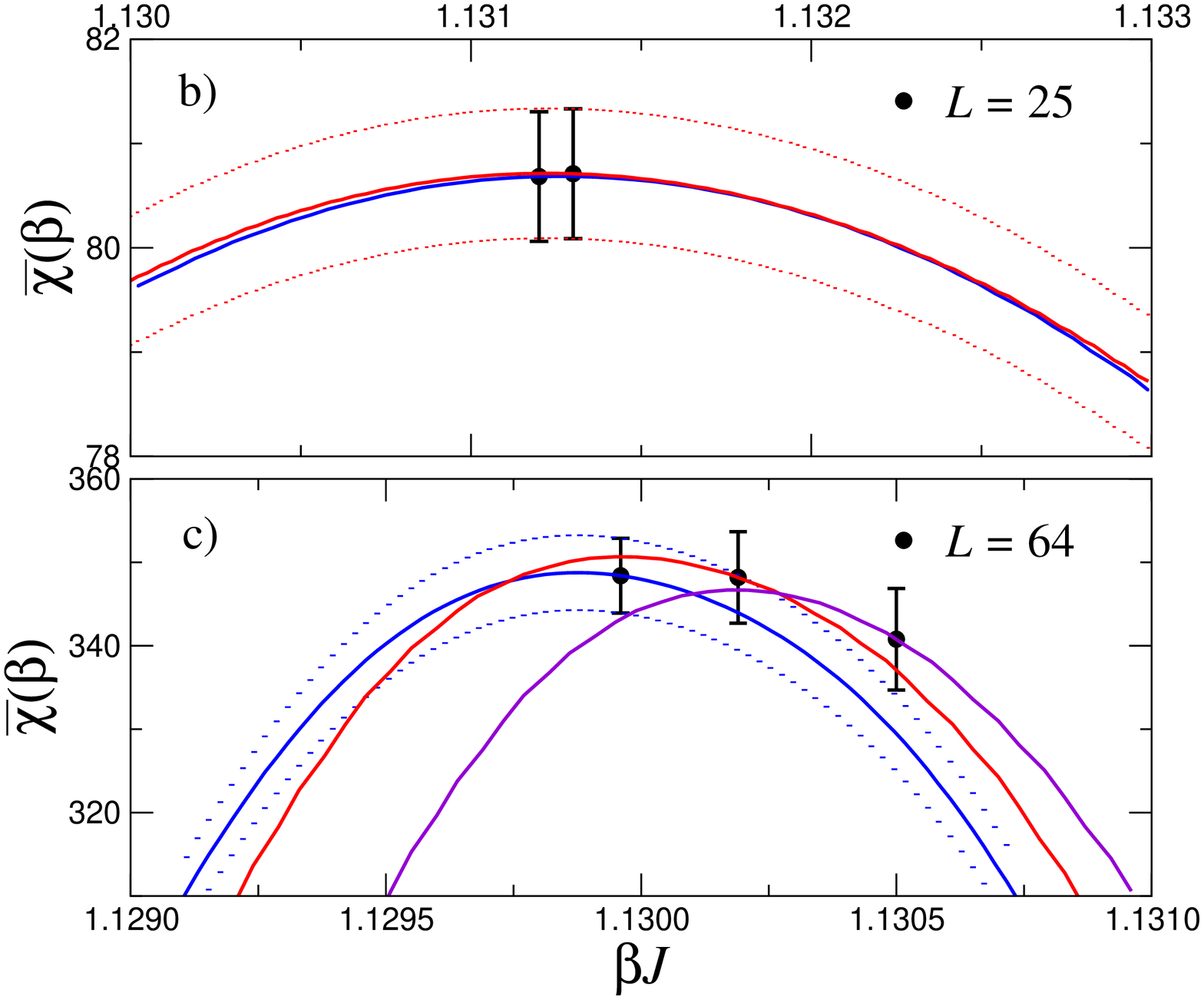}}
        \end{center}\vskip 0cm
        \caption{\small a) Comparison between canonical Swendsen-Wang and multi-bondic 
        algorithms for a pure system ($p=1.0$) of size $L=6$ 
        (histogram reweightings produced
        from simulations at inverse temperatures $\beta J=0.605$
        to $0.655$ are superimposed).
        The insert shows a zoom of the peak location.
        \break
        b) and c) Histogram reweighting of the average susceptibility 
        $\overline\chi(\beta)$ in a disordered system with $p=0.56$
        at different sizes $L=25$ and 64. 
        The maximum is progressively obtained after a few 
        iterations (the next simulation is performed at the temperature
        of the maximum of the histogram reweighting of the current 
        simulation).}
        \label{FigMuca-FigHisto}
\end{figure*}

This is the essential reason for the size limitation
in the first-order regime\footnote{A rough estimate of the time needed by a
single simulation is given by $L^3\times (\# {\rm MCS})\times 1\mu{\rm s}$
for one sample and one temperature.}.
A comparison between the two
algorithms  is illustrated in the case of the pure system for a moderate
size ($L=6$)
in Fig.~\ref{FigMuca-FigHisto}a. The insert shows a zoom of the 
peak in the susceptibility and reveals as expected
that in this first-order regime, the multi-bondic algorithm
provides a better description of the maximum which, since 
being higher is probably
closer to the truth.

In both regimes, the procedure of histogram
reweighting enables us to extrapolate thermodynamic quantities to neighbouring
temperatures. It leads to a better estimate of the transition 
temperature and of the maximum of the susceptibility, 
refining the finite-size estimate at each new size
considered, since the maximum is progressively reached
(Fig.~\ref{FigMuca-FigHisto}b and \ref{FigMuca-FigHisto}c).
The reweighting has to be done for each sample, then the average is
obtained as in Eq.~(\ref{eq2ter}).
For a particular sample, the probability to measure at a given inverse
temperature $\beta$ a microstate $[\sigma]$ with total magnetisation
$M_{[\sigma,J]}=M$ and energy $E_{[\sigma,J]}=E$, is 
$P_\beta(M,E)=(Z_{[J]}(\beta))^{-1}\Omega(M,E)
\ \!{\rm e}^{\beta E}$ where $\Omega(M,E)$ is the degeneracy of
the macrostate. Note that we defined $E$ as minus the energy in order to
deal with a positive quantity. We thus get at a different inverse
temperature $\beta'$
\be
P_{\beta'}(M,E)=(Z_{[J]}(\beta)/Z_{[J]}(\beta'))P_\beta (M,E)
\ \!{\rm e}^{(\beta'-\beta)E},
\ee
where the prefactor $(Z_{[J]}(\beta)/Z_{[J]}(\beta'))$ only depends on the
two temperatures. For any quantity $Q$ depending only on $M_{[\sigma,J]}$ and
$E_{[\sigma,J]}$ the thermal average at the new point $\beta'$ hence follows from
\be
  \langle Q_{[J]}\rangle_{\beta'} = \frac{\sum_{M,E}
         Q(M,E)\ \!P_{\beta}(M,E)
         \ \!{\rm e}^{(\beta'-\beta)E}
          }{\sum_{M,E} P_{\beta}(M,E) \ \!{\rm e}^{(\beta'-\beta)E}}.
\ee
It is well known that the quality of the 
reweighting strongly depends on the number of Monte Carlo iterations,
the larger this number the better the sampling of the configuration space
and thus of the tails of $P_{\beta}$.
Here we have to face up to the disorder average also and a compromise 
between a good disorder statistics and a large temperature scale for the 
reweighting of individual samples has to be found, but we are mainly
interested in the close neighbourhood of the susceptibility maximum, i.e.,
in a small temperature window.

\subsection{Equilibration of the samples and thermal averages}
Before any measurement, each sample has to be in thermal equilibrium at the
simulation temperature. Starting from an arbitrary initial configuration of 
spins, during the initial steps of the simulation process, the system 
explores configurations which are still strongly correlated to the starting
configuration. The typical time
scale over which this ``memory effect'' takes place is measured by the  
autocorrelation time.
The integrated energy autocorrelation time $\tau^e$ 
(one can define more generally an autocorrelation time for any quantity) is
given by 
\be
  \tau_{[J]}^e(\beta)=\frac{1}{2\sigma_e^2}\sum_{i=0}^I\frac{1}{N_{\rm MCS}-I}
  \sum_{j=1}^{N_{\rm MCS}-I}\left(e^j_{[J]} e^{j+i}_{[J]}-\langle e_{[J]}\rangle_\beta^2
  \right),
\label{eq-tau_e}
\ee
where $\sigma_e^2=\langle e_{[J]}^2\rangle_\beta-\langle 
e_{[J]}\rangle_\beta^2$ is the variance, $e^j_{[J]}$ 
is the value of the energy density at iteration $j$ for the realization $[J]$, 
and $I$ is a cutoff (as defined, e.g., by Sokal~\cite{Sokal89}) introduced in 
order to avoid to run a double sum up to $N_{\rm MCS}$, which would
render the estimates very noisy.

\begin{figure*}
        \begin{center}
        \epsfxsize=8.5cm
        \mbox{\epsfbox{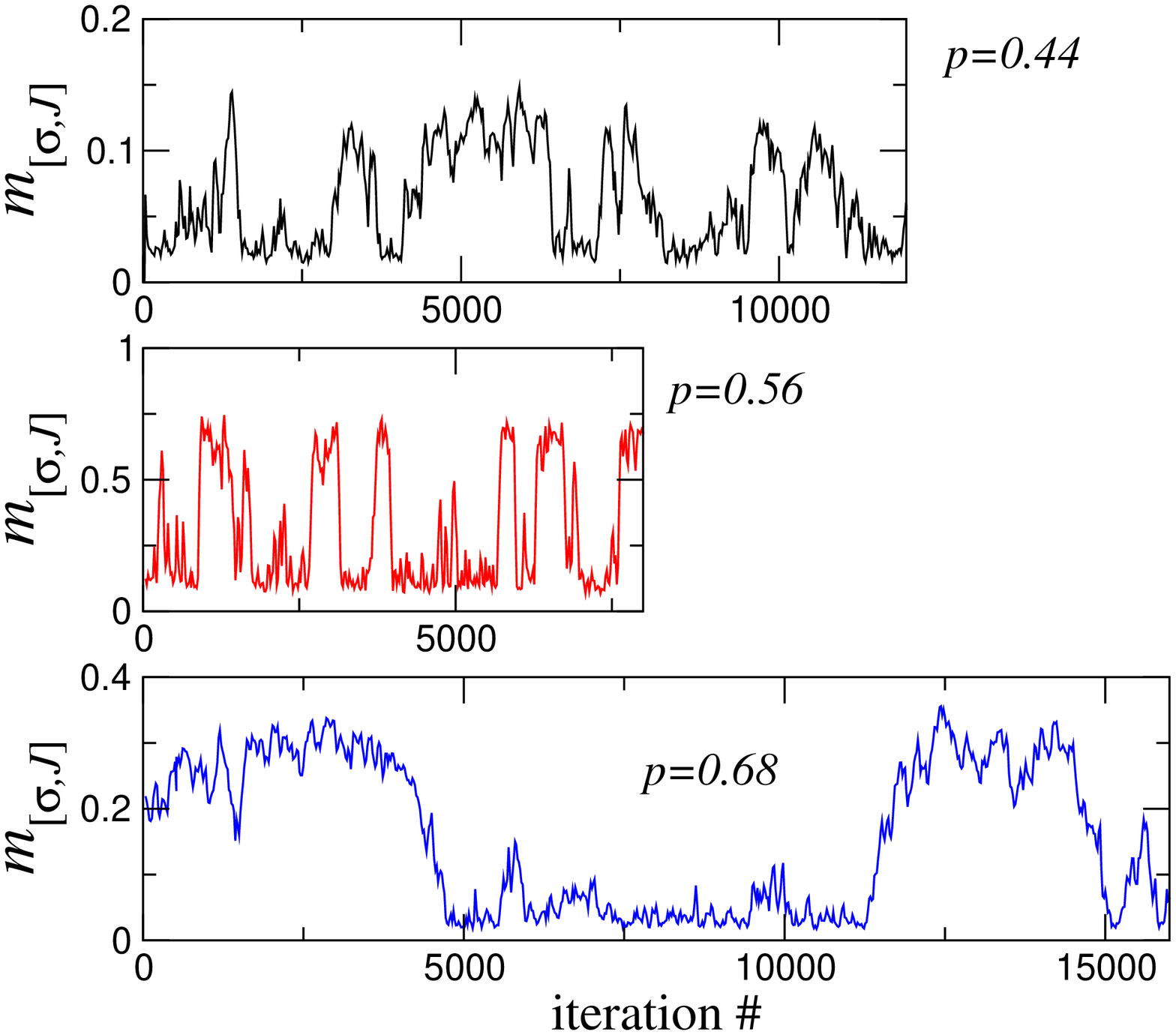}}
        \end{center}\vskip 0cm
        \caption{\small Monte Carlo data of the magnetisation for the
          disorder realization that gave the largest value of 
        $ \chi_{[J]}(\beta_{\rm max}) $ 
        for $p=0.44$ at
          lattice size $L=128$, 
        $p=0.56$ ($L=96$) 
        and $p=0.68$  ($L=50$).}
        \label{Fig11}
\end{figure*}

\begin{figure*}
        \begin{center}
        \epsfxsize=9.5cm
        \mbox{\epsfbox{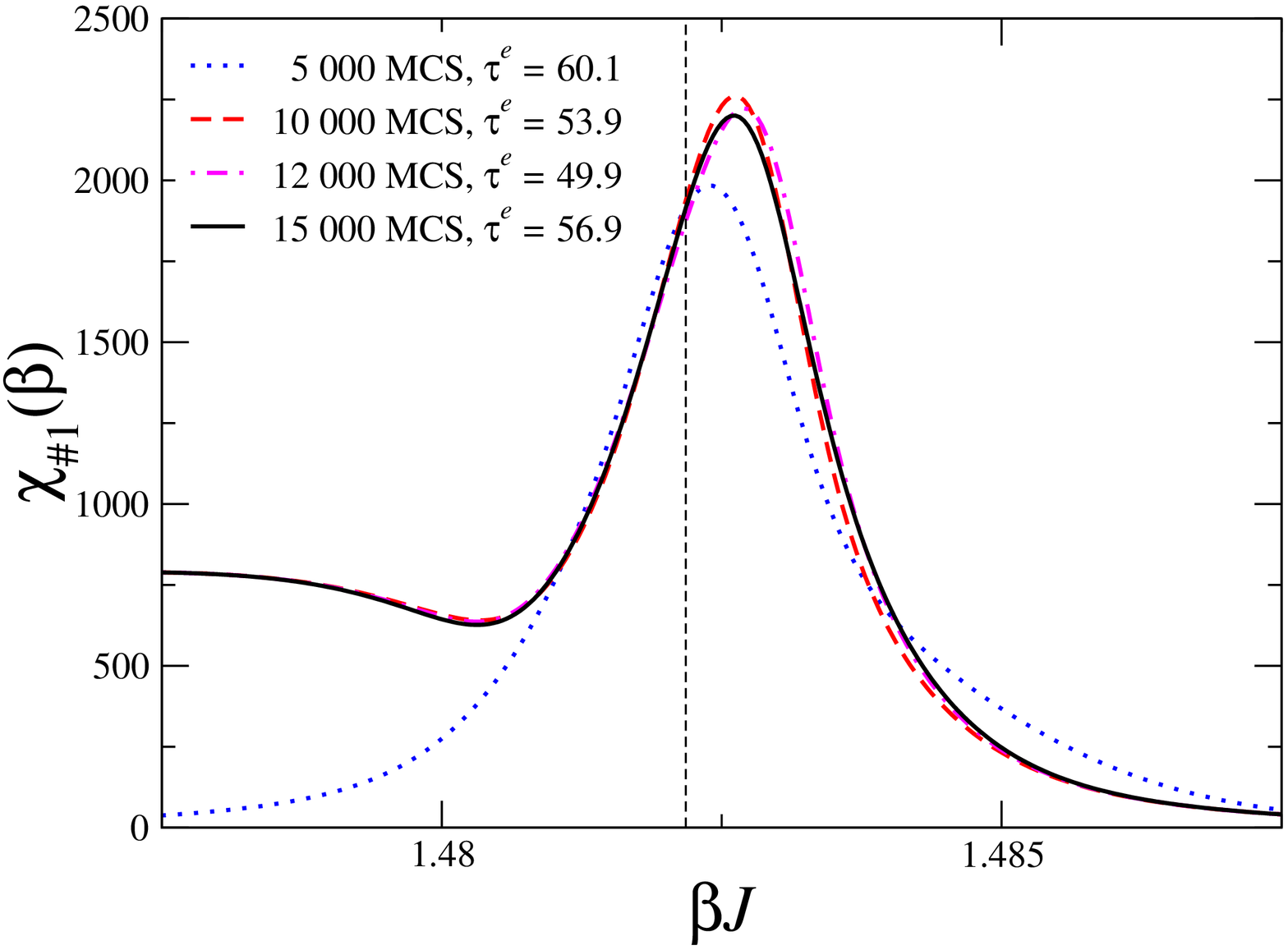}}
        \end{center}\vskip 0cm
        \caption{\small Susceptibility
        for sample $\# 1$ for $p=0.44$ and $L=128$. The 
        different curves show the 
        result of histogram reweighting of simulations close to the maximum
        location after $5\ \!000$, $10\ \!000$, 
        $12\ \!000$, and $15\ \!000$ MCS. We can safely consider that the value
        at the temperature of the maximum of the average susceptibility
        (vertical dashed line) is reliable after $15\ \!000$ MCS.}
        \label{FigHistoChiConfig1}
\end{figure*}

It is worth giving a definition of the errors as computed in this work.
There are two different contributions. Assuming the different realizations
of disorder as completely independent, one has an
error due to randomness on any physical quantity $Q$, defined according to
\begin{equation}
        \Delta_{\rm rdm}\overline Q=\left(\frac{\overline{Q^2}-\overline Q^2}
	{N\{J\}}\right)^{1/2}.
\end{equation}
To the thermal average for each sample is also attached an error which
depends on the autocorrelation time $\tau^Q$, such that the total 
error on a physical quantity is here defined as:
\begin{equation}
\Delta_{\rm tot}\overline Q=\left(\Delta^2_{\rm rdm}\overline Q+
        \frac 1{ N\{J\}}\frac{
        2\overline{\tau^Q}\ \overline{\sigma_Q^2}}{N_{\rm MCS}}\right)^{1/2}.
\end{equation}

For each disorder realization, the preliminary configurations have 
to be discarded and one
usually considers that after 20 times the autocorrelation time $\tau^e$, 
thermal equilibrium is reached. The measurement process can then start and
the thermal average of the physical quantities is considered, in the case
of a single sample, as
satisfying when measurements were done during  typically $10^2\times\tau^e$.
For a quantity $Q$, a satisfactory relative error of the order of 
\be\frac{\Delta_{\rm therm.} \overline Q}{\sqrt{\overline{\sigma_Q^2}}}
=\sqrt{\frac{2\overline{\tau^Q}}{N\{J\}\times N_{\rm MCS}}}
\simeq 10^{-2}
\ee 
indeed requires typically  $N\{J\}\times N_{\rm MCS}\simeq 10^4\overline{\tau^Q}$.
Since we also need a large number of disorder realizations in order to
minimise $\Delta^2_{\rm rdm}\overline Q$, typically 
$N\{J\}\simeq 10^2-10^4$, each sample requires a ``production'' process
during $N_{\rm MCS}\simeq (10^0-10^2)\overline{\tau^Q}$. In this paper, we choose 
to work at the upper limit with $N_{\rm MCS}> 10^2\tau^e$
(since there is a single dynamics in the algorithm, the time scale is
usually measured through the energy autocorrelation time) and
$N\{J\}> 10^3$.

Examples of times series of the magnetisation are shown in Fig.~\ref{Fig11}
for particular samples (those which contribute the most to the average
susceptibility) at the three
largest sizes studied\footnote{The main illustrations are shown in the
worst cases, i.e., for the largest systems at each dilution.} 
at dilutions $p=0.44$, $0.56$, and $0.64$ in the
second-order regime. The simulation temperature is extremely close to
the transition temperature and tunnelling between ordered and disordered
phases guarantees a reliable thermal average.

Another test of thermal equilibration is given by 
the influence of the number of MCS which
are taken into account in the evaluation of thermal averages.
An example is shown in Fig.~\ref{FigHistoChiConfig1} where the histogram
reweightings of the susceptibility, as obtained with different MCS $\#$'s,
are shown for a typical sample (the first sample, $\#\ \! 1$, 
is in fact supposed
to be typical). Although quite different
far from the simulation temperature (which is close to the maximum) the 
different curves are in a satisfying agreement at the temperature 
$\beta_{\rm max}$ of the maximum of the average susceptibility, shown by
a vertical dashed line. The criterion $\#\ \!$MCS~$\ge 250\times\tau^e$ is
safely satisfied for the larger number of iterations\footnote{We also
note that something happened between 5000 and 10000 MCS, since the 
shape of $\chi_{\# 1}$ at high temperatures becomes unphysical. This is an 
illustration of the finite window of confidence of the histogram reweighting
procedure.}.

\subsection{Properties of disorder averages} 
\begin{figure}[h]
        \epsfxsize=10cm
        \begin{center}
        \mbox{\epsfbox{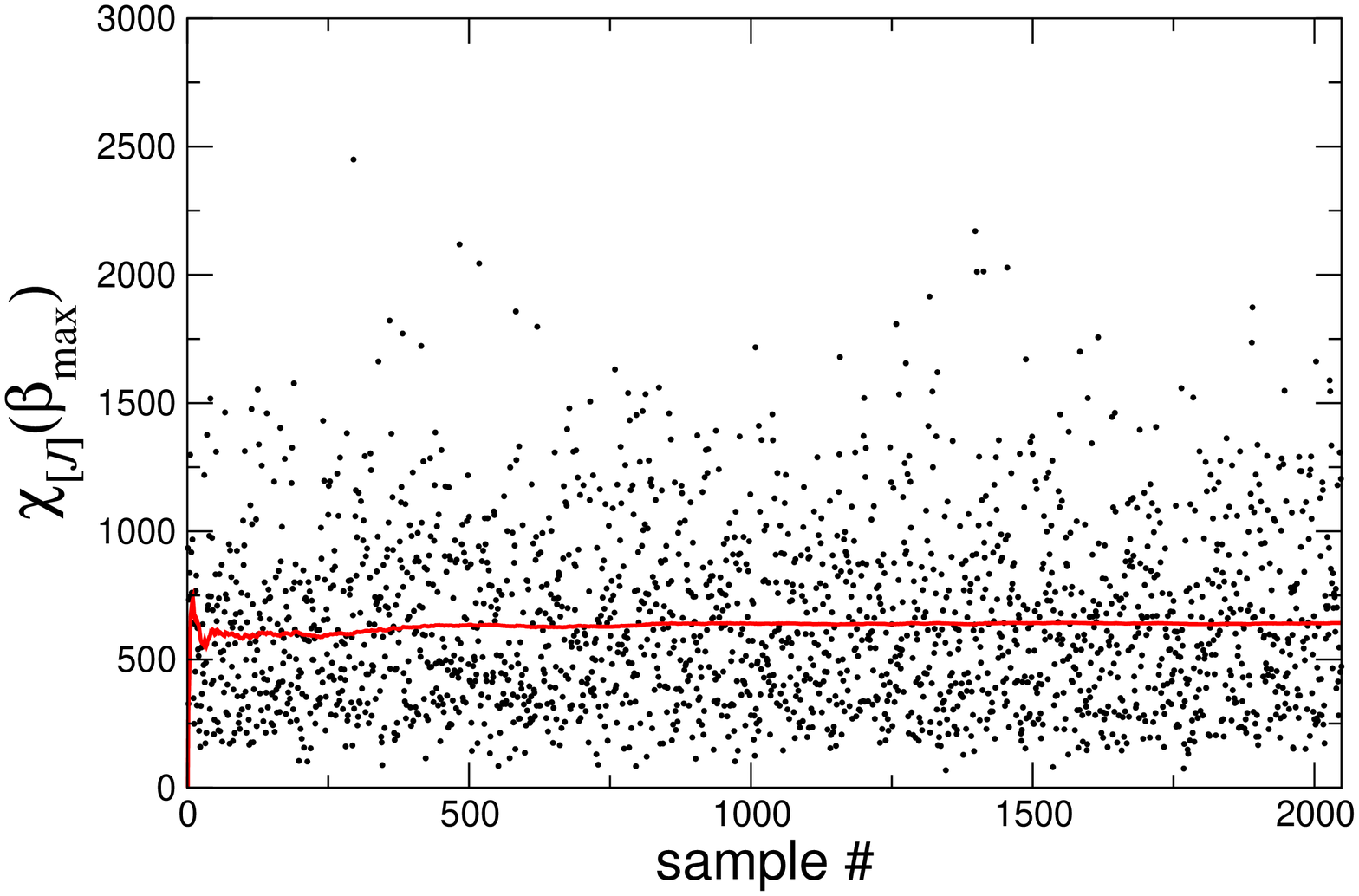}}
        \end{center}\vskip 0cm
        \caption{\small Different values of 
        $ \chi_{[J]}(\beta_{\rm max})$ (over the samples) 
        of susceptibility at $p=0.56$, $L=96$ (the simulation is performed at 
        the temperature of the maximum of the average susceptibility).
        The running average over the samples is shown by the solid line.}
        \label{FigConvChi96}
\end{figure}

For different samples, corresponding to distinct disorder realizations,
the susceptibility $ \chi_{[J]}(\beta) $ 
at thermal equilibrium may have very different
values (see Fig.~\ref{FigConvChi96} where the running average over the samples
is also shown and remains stable after a few hundreds of realizations).

\begin{figure*}
        \epsfxsize=12.0cm
        \begin{center}
        \mbox{\epsfbox{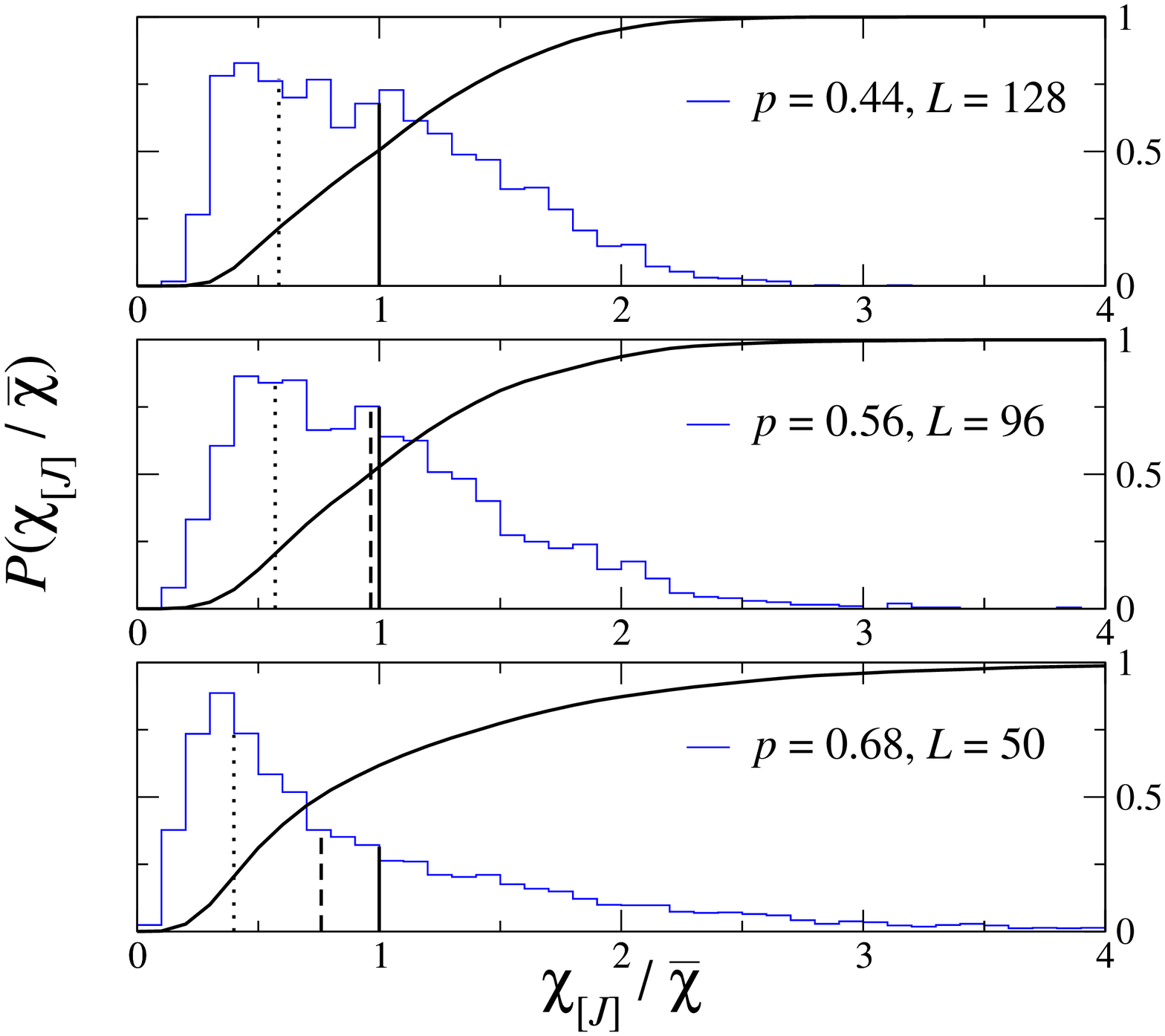}}
        \end{center}\vskip 0cm
        \caption{\small Probability distribution of the susceptibility 
        $ \chi_{[J]}(\beta_{\rm max}) $
          for the bond concentrations $p=0.44$, $0.56$, and $0.68$ 
          for the largest lattice size in each case. The full curve represents the
        integrated distribution. At each dilution, a full 
        vertical line shows the 
        location of the average susceptibility, a dashed line shows the median
        and a dotted line shows the average over the events which are smaller
        than the median.}
        \label{Fig12}
\end{figure*}

We paid attention to average the data over a sufficiently large number of
disorder realizations (typically 2000 to 5000) to ensure
reliable estimates of non-self-averaging quantities~\cite{Derrida84}.
Averaging over a too
small number of random configurations leads to typical (i.e. most probable)
values instead of average ones. Indeed, as can be seen in Fig.~\ref{Fig12}, the
probability distribution of $ \chi_{[J]} $ 
(plotted at the temperature $\beta_{\rm max}$ where the
average susceptibility is maximum) presents a long tail of rare events 
with large values of the susceptibility. These samples have a large 
contribution to the average, shifted far from the most probable value. 
The larger the value of $p$, 
the longer the tail.
Scanning the regime close to the first-order transition thus requires large
numbers of samples to explore efficiently the configuration space,
so the simulations
were limited to $L=50$ at $p=0.68$ while we made the calculation up to
$L=128$ at $p=0.44$.
In the example of Fig.~\ref{Fig11}, the thermodynamic quantities have been
averaged over $3500$ disorder realizations for $p=0.44$ at lattice size $L=128$,
$2048$ for $p=0.56$, $L=96$, and $5000$ disorder realizations 
for $p=0.68$, $L=50$.

\begin{figure}[ht]
        \epsfxsize=10.0cm
        \begin{center}
        \mbox{\epsfbox{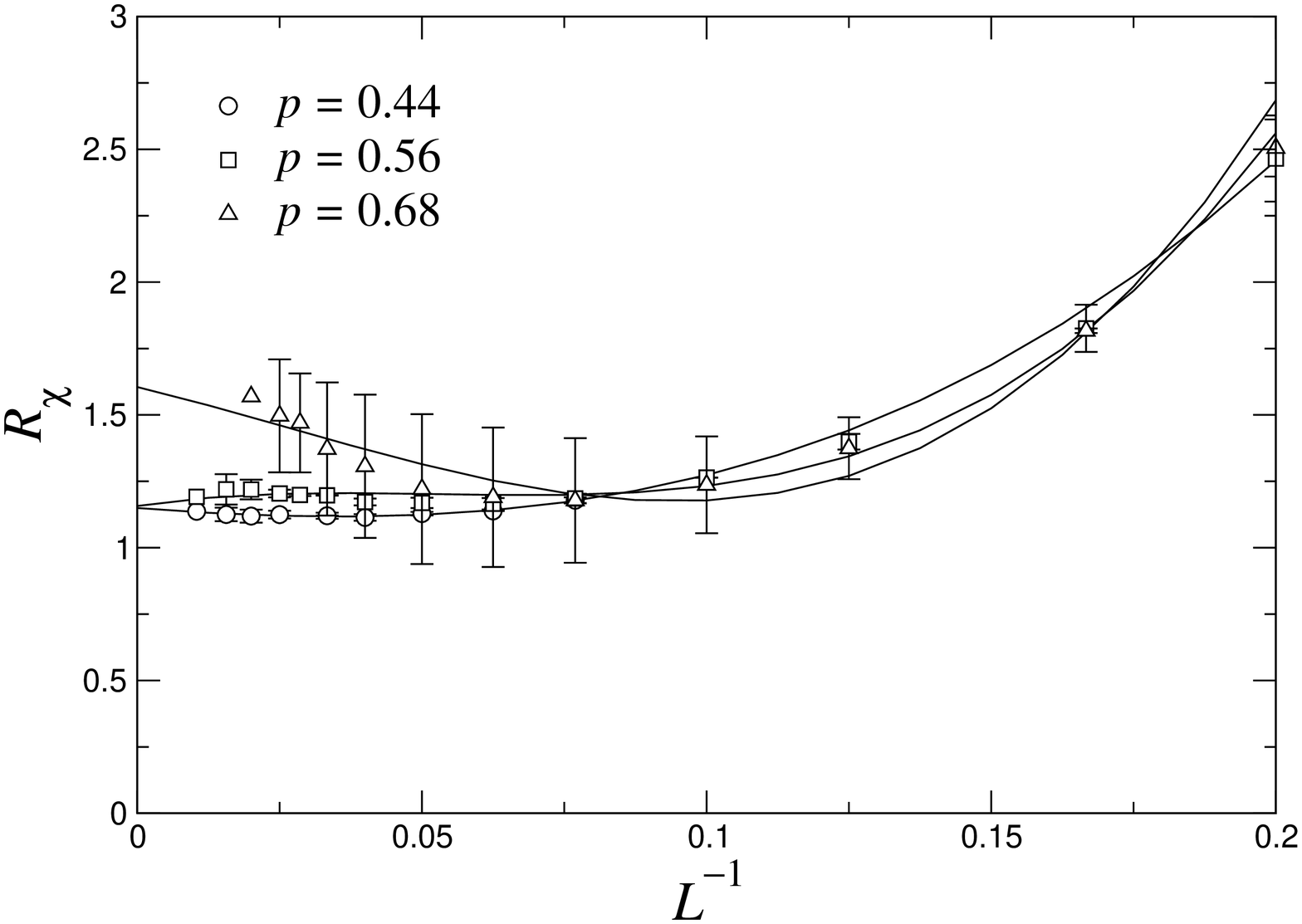}}
        \end{center}\vskip 0cm
        \caption{\small Normalised squared width of the susceptibility, $R_\chi$,
        plotted against the inverse lattice size for the three dilutions
        $p=0.44$, 0.56, and 0.68. The solid lines are polynomial fits used as 
        guides for the eyes. Note that $\chi$ is apparently less and less
        self-averaging as $p$ increases.}
        \label{FigR_chi}
\end{figure}

Self-averaging properties are quantified through the normalised squared
width, for example in the case of the susceptibility,
$R_\chi=\sigma_\chi^2 (L)/\overline\chi^2$, where $\sigma_\chi^2=
\overline{\chi^2}-\overline\chi^2$. For a self-averaging quantity, say 
$Q$, the probability distribution, albeit not truly Gaussian, may be
considered so in first approximation close to the peak, and 
$P_\beta(Q)\simeq (2\pi\sigma_Q^2)^{-1/2}{\rm e}^{-(Q
-\overline{\langle Q\rangle})^2/2\sigma_Q^2}$ evolves towards a sharp
peak in the thermodynamic limit, $P_\beta(Q)\to_{L\to\infty}\delta(Q
-\overline{\langle Q\rangle})$. The probability of the average event
$\overline{\langle Q\rangle}$ goes to 1 and 
the normalised squared width evolves towards zero in the thermodynamic
limit while it keeps a finite value for a non-self-averaging quantity,
as shown in the case of the susceptibility in Fig.~\ref{FigR_chi}.
The observation of a longer tail in the probability distribution of the
$\chi$ values when $p$ increases is expressed in Fig.~\ref{FigR_chi} by
the fact that $\chi$ becomes less and less self-averaging 
when $p$ increases.

\begin{figure*}
        \epsfxsize=10.0cm
        \begin{center}
        \mbox{\epsfbox{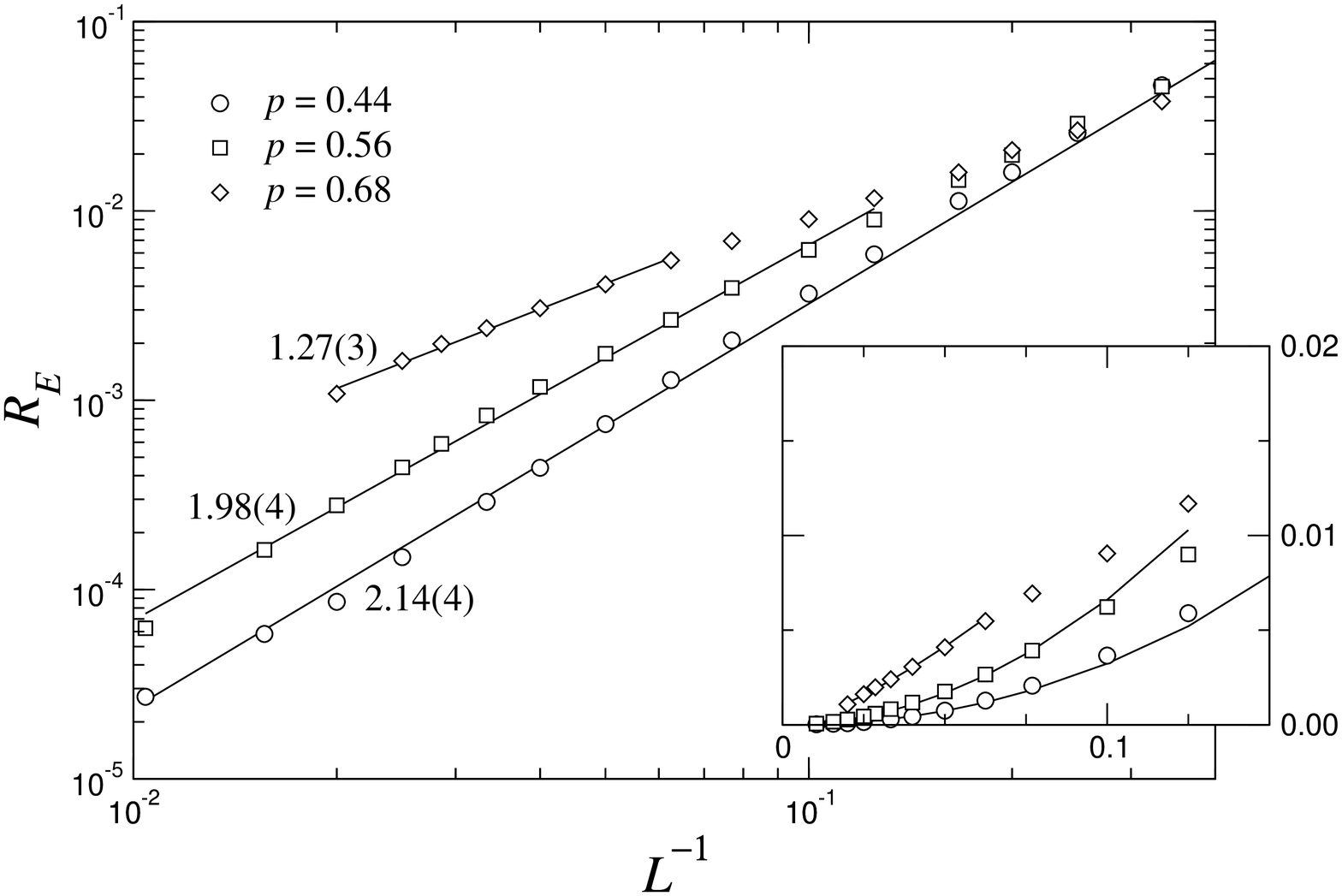}}
        \end{center}\vskip 0cm
        \caption{\small Normalised squared width of the energy, $R_{E}$
        plotted on a log-log scale against the inverse lattice size for the 
	three dilutions
        $p=0.44$, 0.56, and 0.68. Power-law fits have been performed
        and corresponding exponents printed by the curve. The insert presents
        the same data plotted on a linear scale.}
        \label{FigR_E}
\end{figure*}

In contradistinction to the magnetic susceptibility, the energy seems to be
weakly self-averaging in the range of lattice sizes that we studied 
as seen in Fig.~\ref{FigR_E}. The associated exponent depends on the 
concentration of bonds $p$. This concentration dependence may be 
effective and due to corrections generated by other fixed points (see below).

In Table~\ref{TabChi1}, 
the influence of the number of MCS is shown for
typical samples, but also for the average susceptibility. Although the 
variations for a given sample and from sample to sample are important, the
average seems stable with our choice of number of iterations (the largest),
and also the autocorrelation time (for the average) is stable.

{\small
\begin{table*}
\caption{\small Evolution of the susceptibility with the number of
        Monte Carlo sweeps per spin for different samples, 
        $ \chi_{[J]} $ and the average 
        value (with 2048 samples) at $p=0.56$, $L=96$. The data are given at 
        the maximum location of the average susceptibility, $\beta_{\rm max}$.
        The last column gives the number of independent measurements per 
        sample.
        \label{TabChi1}
        }
\bcenter\begin{tabular}{rrrrrrlrc}
\noalign{\vspace{3mm}}
        \hline\noalign{\vspace{0.1pt}}
        \hline\noalign{\vspace{0.1pt}}
        \hline
        $N_{\rm MCS}$& $ \chi_{\#1} $ & 
        $ \chi_{\#2} $ 
        & $ \chi_{\#3} $ & 
        $ \chi_{\#4} $ 
        & $ \chi_{\#5} $ & 
        $\overline\chi_{\rm max}$ 
        & $\tau^e(\beta_{\rm max})$ & meas. $/$ sample\\
        \hline
 5000 & 994 & 404 & 611 & 682 & 1803 & 617(8) &  95.1 & $\simeq \,~50$ \\
10000 & 952 & 390 & 698 & 614 & 1574 & 634(8) & 107.4 & $\simeq \,~90$ \\
15000 & 1010& 356 & 680 & 819 & 1398 & 638(8) & 111.7 & $\simeq 130$ \\
20000 & 939 & 351 & 689 & 851 & 1320 & 641(7) & 114.0 & $\simeq 175$ \\
25000 & 911 & 327 & 675 & 848 & 1308 & 643(8) & 115.3 & $\simeq 200$ \\
30000 & 934 & 327 & 733 & 837 & 1297 & 643(8) & 116.9 & $\simeq 250$ \\
        \hline\noalign{\vspace{0.1pt}}
        \hline\noalign{\vspace{0.1pt}}
        \hline  
\end{tabular}\\[0.5cm]
        \ecenter\end{table*}
}

In Fig.~\ref{Fig12},  a full vertical 
line points out the location of the average
susceptibility $\overline\chi_{\rm max}$. 
In order to give a comparison, the median value $\chi_{\rm med}$,
defined as the value of $ \chi_{[J]} $ 
where the integrated probability takes 
the value $50\%$, is shown as the dashed line. The more it differs from the
average, the more asymmetric is the probability distribution.
This is more pronounced when $p$ increases.
We also notice that the maximum of the probability distribution (the typical
samples) corresponds 
to smaller susceptibilities. For a given number of disorder
realizations, this peak is better described than the tail at larger 
susceptibilities, so we also define (shown as dotted lines) an average over 
the samples smaller than the median susceptibility, that we denote
$\chi_{50\%}$, 
\be
\chi_{50\%}=
2\int_0^{\chi_{\rm med}}  \chi_{[J]} 
P_\beta( \chi_{[J]} )\ \!{d} \chi_{[J]},\quad\quad
\int_0^{\chi_{\rm med}} P_\beta( \chi_{[J]} )\ \!{d} \chi_{[J]}={1\over 2},
\label{chi50}
\ee
where the factor $2$ normalises the truncated distribution.
In the particular case of the probability distributions observed
here, i.e.\ with a sharp initial increase, a peak located at small events and a 
long tail at large values of the variable\footnote{This shape of probability
distribution is very different than in the case of the 3D dilute Ising 
model~\cite{BCBJ04}.
}, 
this definition empirically gives a sensitive measure of
the typical or most probable value. 
We shall refer to this quantity when typical behaviour will be concerned.

\section{Qualitative description of the transition\label{sec4}}
Before performing a quantitative analysis of the transition, it is interesting
to study in some detail why the probability distributions
have significantly different shapes when $p$ varies, and which type of
sample can be considered as a typical one, or which one 
corresponds to a rare event with quite a large or very small
susceptibility. Here we shall focus on
the second-order regime and in particular on $p=0.44$ for the largest simulated
size, $L=128$, for which the probability distribution of $\chi_{[J]}$ at
$\beta_{\rm max}$ can be inspected in Fig.~\ref{Fig12}.

\begin{table}
\caption{\small Relative variations of the peak height 
        $\Delta\chi_{[J]}/\bar\chi_{\rm max}$ and peak location
        $\Delta \beta_{[J],{\rm max}}/\beta_{\rm max}$
        for a few samples, chosen among the rare and the typical 
        samples at $p=0.44$, $L=128$.
        For reference, the values of the average are given by
        $\beta_{\rm max}J=1.4820$, 
        $\overline\chi_{\rm max}=1450$. The asterisks $(*)$ mark those
        samples that are discussed in detail in Figs.~\ref{FigRareHigh-0.44-128_1} --
        \ref{FigRareLow-0.44-128_2}.}
        \label{TabChiRareAndTypical}
\bcenter\begin{tabular}{clrlrll}
        \noalign{\vspace{3mm}}
        \hline\noalign{\vspace{0.4pt}}
        \hline\noalign{\vspace{0.4pt}}
        \hline
        type  & sample $\#$ & $\chi_{[J],{\rm max}}$ 
        & $\beta_{[J],{\rm max}}J$ & 
        $\Delta\chi_{[J]}/\bar\chi_{\rm max}$ & 
        $\Delta \beta_{[J],{\rm max}}/\beta_{\rm max}  $ \\
        \hline
               & 0035 (*) & 5253 & 1.4823 & $262.3~\%$ & $\,~~0.02~\%$ \\
rare           & 0438     & 3862 & 1.4822 & $166.3~\%$  & $\,~~0.013~\%$ \\
(large $\chi$) & 1135     & 3825 & 1.4821 & $163.8~\%$  & $\,~~0.007~\%$\\
               & 3302     & 4314 & 1.4823 & $197.5~\%$  & $\,~~0.02~\%$ \\
        \hline
               & 0006     & 1550 & 1.4831 & $6.9~\%$   & $\,~~0.07~\%$ \\
typical        & 0008 (*) & 2792 & 1.4810 & $92.5~\%$   & $-0.07~\%$ \\
(around peak)  & 0021     & 1473 & 1.4819 & $1.6~\%$   & $-0.007~\%$ \\
               & 0039     & 2345 & 1.4817 & $61.7~\%$   & $-0.02~\%$ \\
        \hline
               & 0373     & 946  & 1.4852 & $-34.7~\%$   & $\,~~0.22~\%$ \\
rare           & 1492 (*) & 286  & 1.4830 & $-80.3~\%$ & $\,~~0.07~\%$ \\
(small $\chi$) & 1967     & 1063 & 1.4847 & $-26.7~\%$    & $\,~~0.2~\%$ \\
               & 2294     & 769  & 1.4853 & $-46.9~\%$    & $\,~~0.2~\%$ \\
        \hline\noalign{\vspace{0.4pt}}
        \hline\noalign{\vspace{0.4pt}}
        \hline
\end{tabular}\\[0.5cm]
\ecenter\end{table}

\begin{figure}[h]
        \epsfxsize=11.5cm
        \begin{center}
        \mbox{\epsfbox{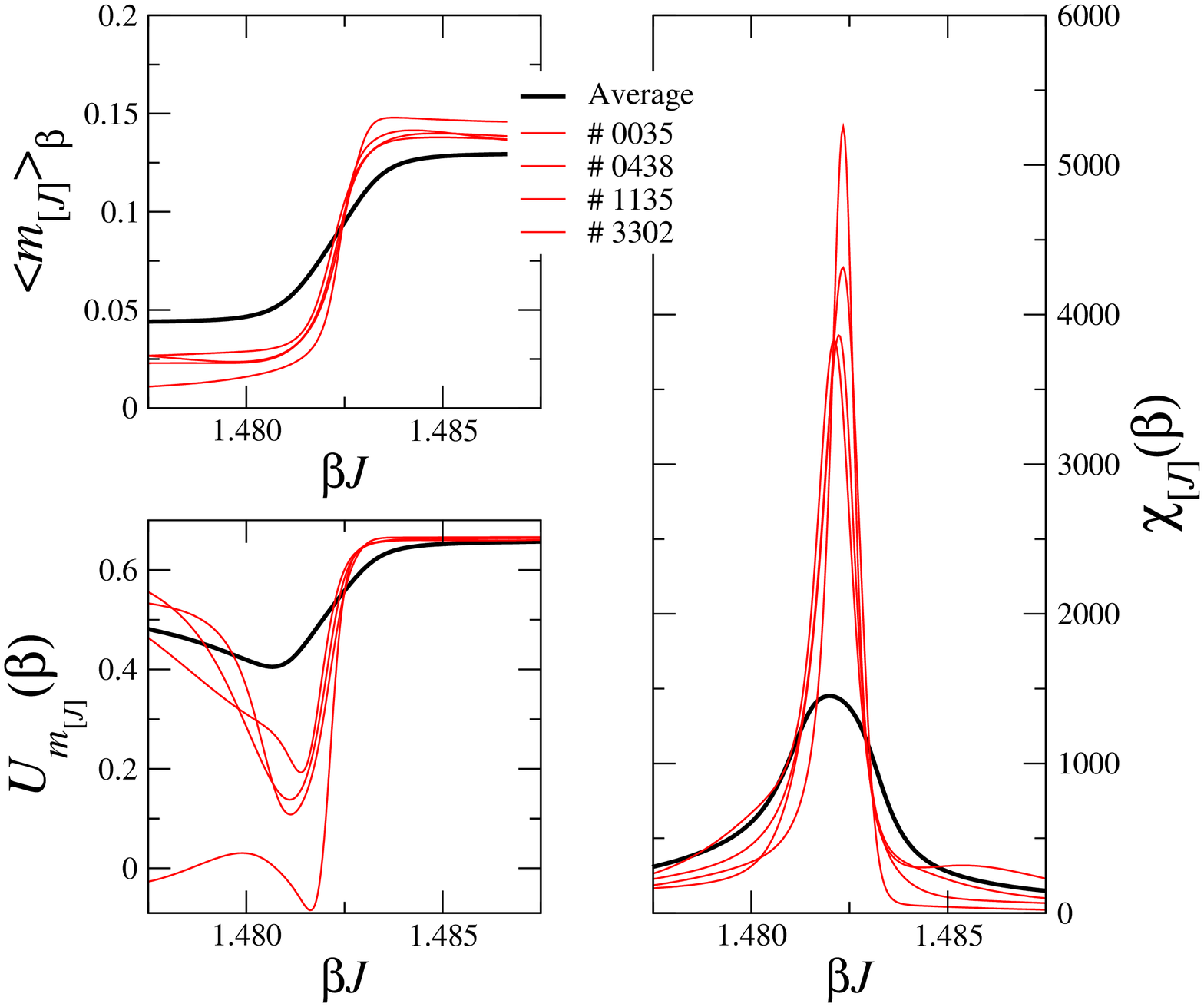}}
        \end{center}\vskip 0cm
        \caption{\small Examples of {\em rare\/} events for $p=0.44$ and $L=128$.
        with {\em large\/} values of $ \chi_{[J]}$.
 	The thick lines show the averages over all realizations.
         }
        \label{FigHistoChiRare-0.44-128}
\end{figure}

Each sample displays its own maximum and due to the fluctuations over disorder, 
the temperature $\beta_{[J],{\rm max}}$ where it occurs varies from sample to sample.
In Table~\ref{TabChiRareAndTypical}, we quote for a few rare and typical samples
the values of $\chi_{[J]_{\rm max}}$ and $\beta_{[J],{\rm max}}$, the maximum
of the sample susceptibility and the corresponding inverse temperature
(see also Figs.~\ref{FigHistoChiRare-0.44-128} to \ref{FigHistoChiFoot-0.44-128}).
The relative variations of these numbers with respect to their average values at
$\beta_{\rm max}$~: $\Delta\chi_{[J]}/\bar\chi_{\rm max}=[\chi_{[J],{\rm max}}
-\bar\chi_{\rm max}]/\bar\chi_{\rm max}$, and $\Delta\beta_{[J]}/\beta_{\rm max}
=[\beta_{[J],{\rm max}}-\beta_{\rm max}]/\beta_{\rm max}$ are also collected in
Table~\ref{TabChiRareAndTypical}. It turns out that rare events with large
susceptibility do also display a very small shift of the temperature
$\beta_{[J],{\rm max}}$ with respect to the average. Other events have a smaller
susceptibility at $\beta_{\rm max}$ both because their maximum $\chi_{[J],{\rm max}}$
is smaller but also because of a larger shift of the temperature
$\beta_{[J],{\rm max}}$ where this maximum occurs.
A few examples of {\em rare\/} events corresponding to {\em large\/} values
of $ \chi_{[J]} $ are shown in Fig.~\ref{FigHistoChiRare-0.44-128}.
{\em Rare\/} events corresponding to {\em small\/} values of $ \chi_{[J]} $
are presented in Fig.~\ref{FigHistoChiFoot-0.44-128}. They have a very small
contribution to the phase transition so in the following, we will refer only to
events with large values of $\chi_{[J]}$ when mentioning rare events.
In Fig.~\ref{FigHistoChiTypical-0.44-128}, the same is done for {\em typical\/}
events, i.e., those for which the values of $\chi_{[J]}$ are in the peak of the
distribution. The scales of both axis are the same in the three figures in order
to facilitate the comparison.

\begin{figure}[t]
        \epsfxsize=11.5cm
        \begin{center}
        \mbox{\epsfbox{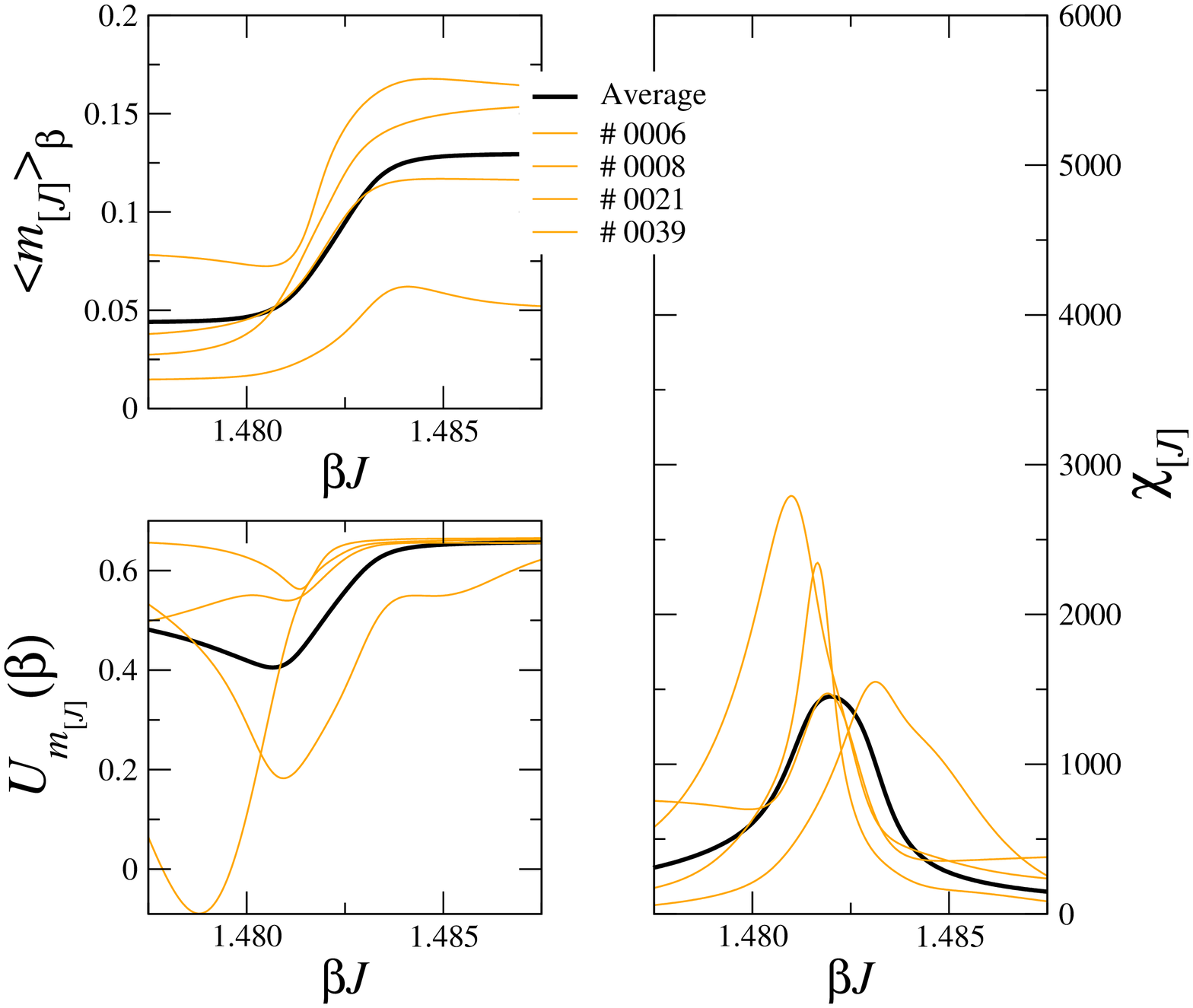}}
        \end{center}\vskip 0cm
        \caption{\small Examples of {\em typical\/} events for the same parameters as
        in Fig.~\ref{FigHistoChiRare-0.44-128}. The thick lines show the averages
        over all realizations.
         }
        \label{FigHistoChiTypical-0.44-128}
\efig

\bfig
         \epsfxsize=11.5cm
         \begin{center}
         \mbox{\epsfbox{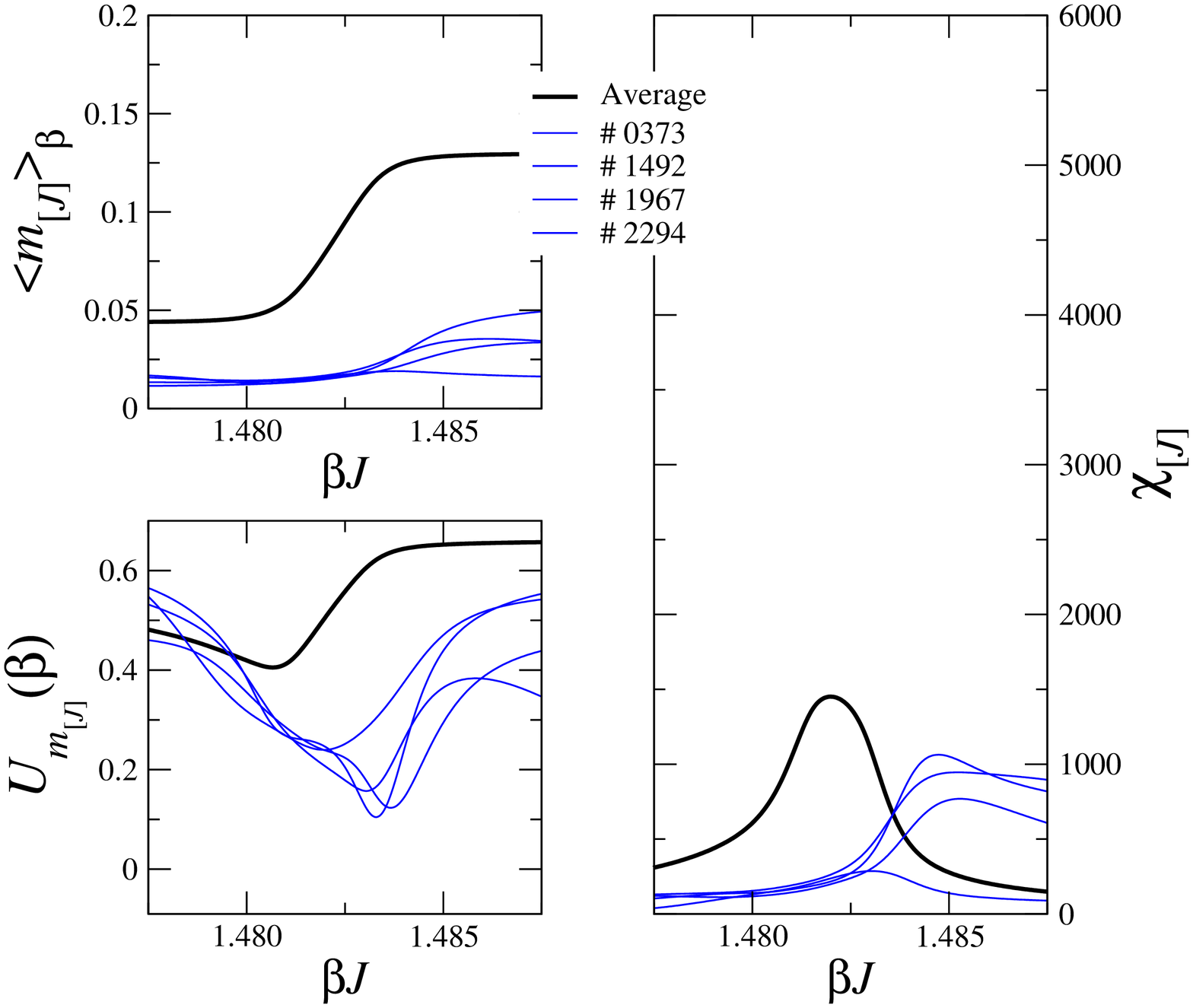}}
         \end{center}\vskip 0cm
         \caption{\small Examples of {\em rare\/} events for the same parameters as
         in Fig.~\ref{FigHistoChiRare-0.44-128}
 	with $\chi_{[J]} $ at the foot of the probability distribution. The thick
 	lines show the averages over all realizations.
          }
         \label{FigHistoChiFoot-0.44-128}
\efig

\bfig
        \epsfxsize=10.0cm
        \begin{center}
        \mbox{\epsfbox{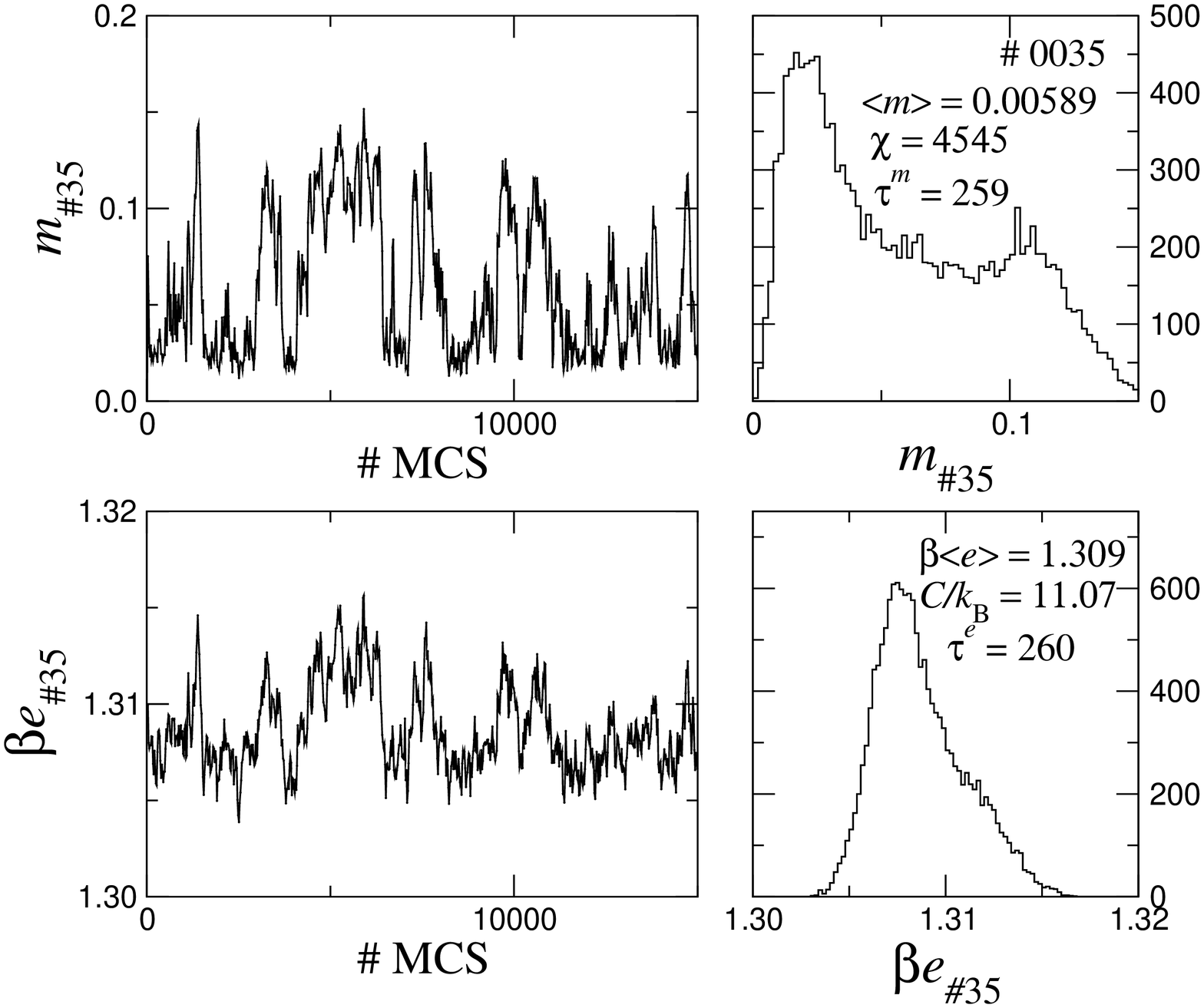}}
        \end{center}\vskip 0cm
        \caption{\small Time series of the magnetisation and the energy density
        and corresponding probability distributions for a {\em rare\/} event
        ($\# 35$) with {\em large\/} $\chi_{[J]}$ ($p=0.44$, $L=128$, 
        simulation at inverse temperature $\beta J=1.48218$).}
        \label{FigRareHigh-0.44-128_1}
\efig
\bfig
        \epsfxsize=10.0cm
        \begin{center}
        \mbox{\epsfbox{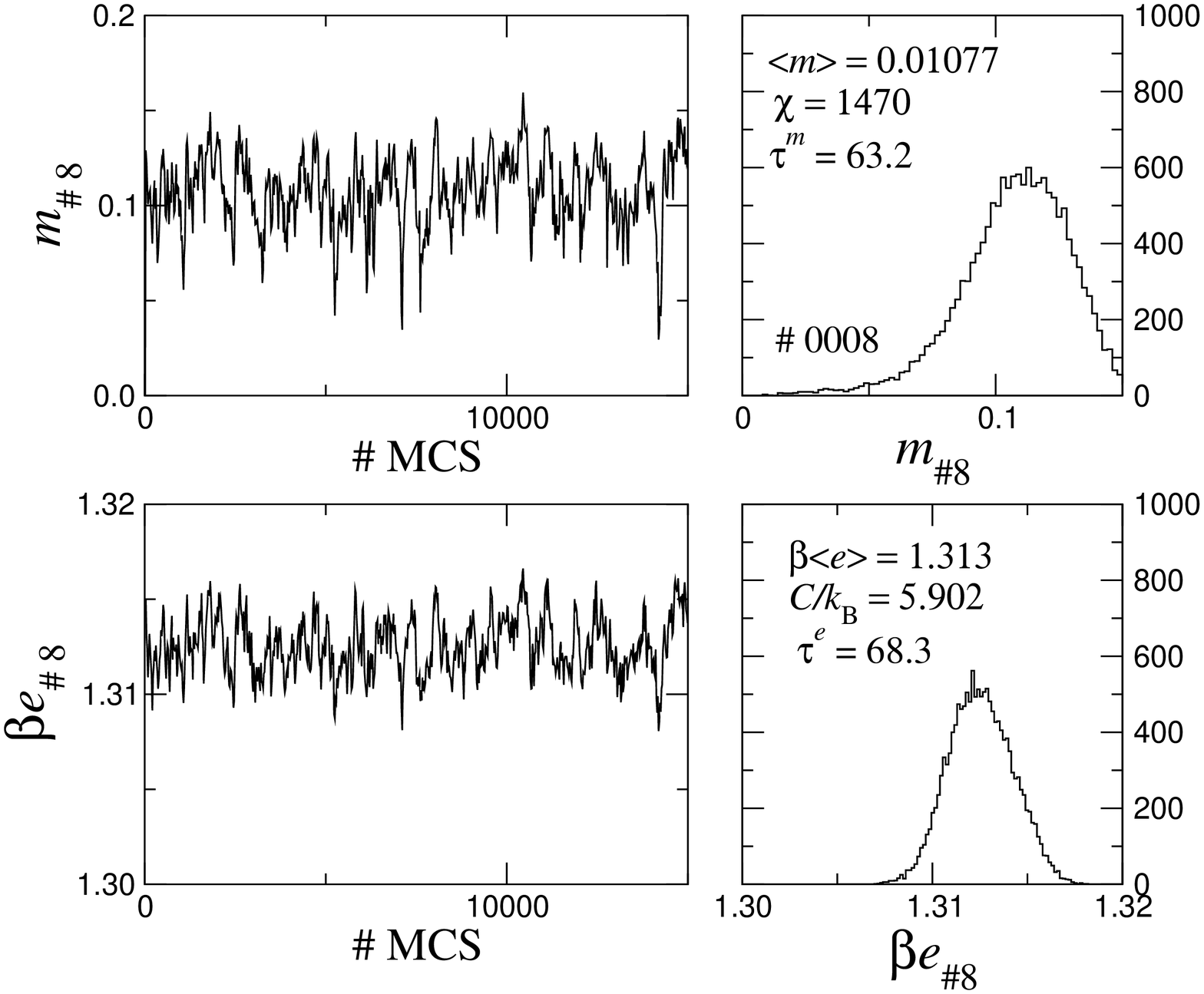}}
        \end{center}\vskip 0cm
        \caption{\small Time series of the magnetisation and the energy density
        and corresponding probability distributions for a {\em typical} event
        ($\# 8$) with $\chi_{[J]}$ in the vicinity of the most probable value
        ($p=0.44$, $L=128$, simulation at inverse temperature $\beta J=1.48218$).}
        \label{FigTypical-0.44-128_2}
\efig
\bfig
        \epsfxsize=10.0cm
        \begin{center}
        \mbox{\epsfbox{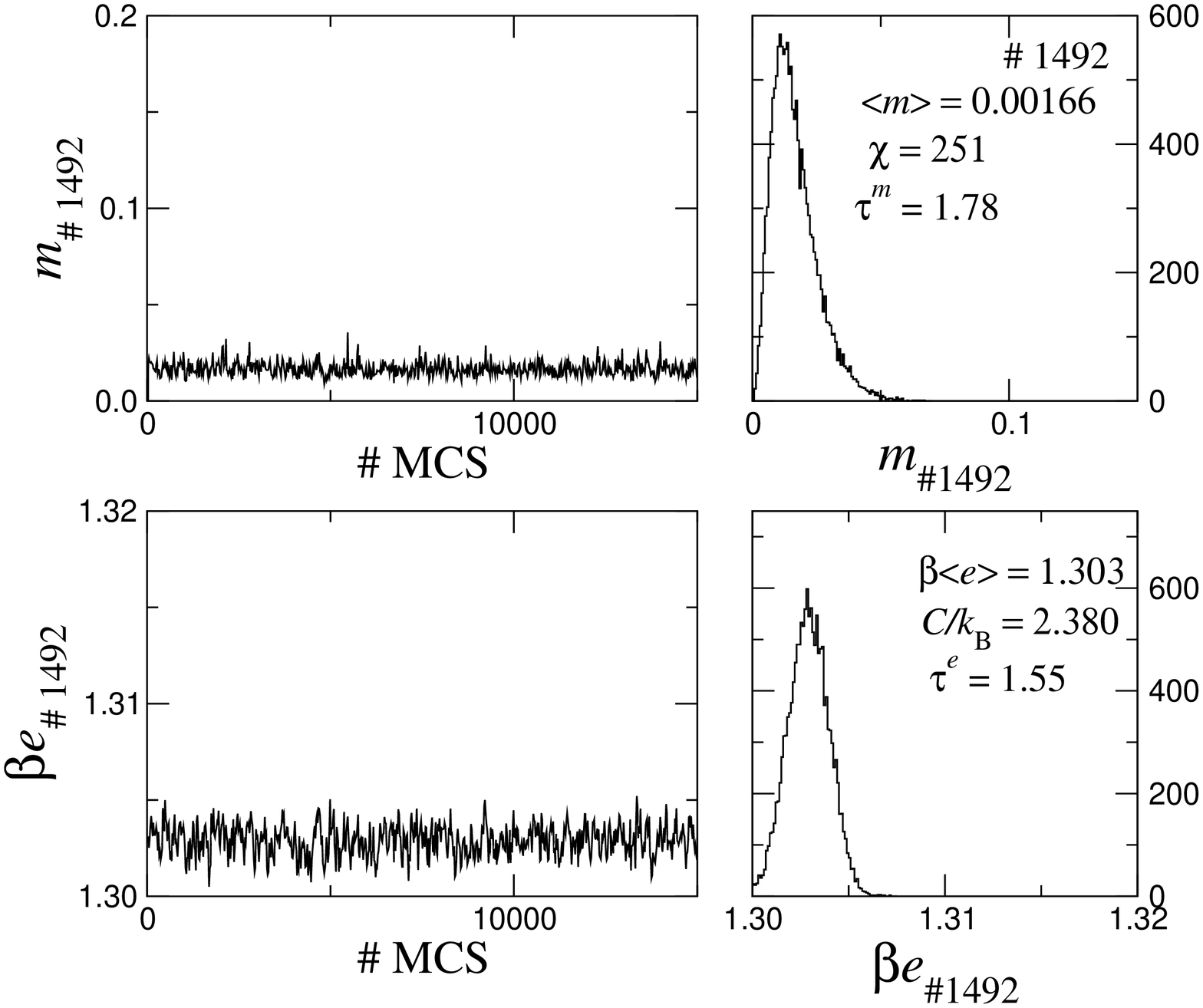}}
        \end{center}\vskip 0cm
        \caption{\small Time series of the magnetisation and the energy density
        and corresponding probability distributions for a {\em rare\/} event 
        ($\# 1492$) with very
        {\em small\/} $\chi_{[J]}$, which looks similar to typical events 
        ($p=0.44$, $L=128$, simulation at inverse
        temperature $\beta J=1.48218$).}
        \label{FigRareLow-0.44-128_2}
\efig

In Figs.~\ref{FigRareHigh-0.44-128_1}-\ref{FigRareLow-0.44-128_2}, 
we can follow the fluctuations of the magnetisation
during the thermalisation process (after equilibration) 
for three different samples. Configuration $\# 35$
(Fig.~\ref{FigRareHigh-0.44-128_1}) corresponds to a
rare event, with the definition given above, while the sample
$\# 8$ (Fig.~\ref{FigTypical-0.44-128_2}) 
is a typical one. The last sample, $\# 1492$ 
(Fig.~\ref{FigRareLow-0.44-128_2}), is an example of
a realization of disorder which leads to a very small susceptibility peak.
These figures also present the magnetisation and energy probability 
distributions. The rare event (Fig.~\ref{FigRareHigh-0.44-128_1})
displays a double-peak structure in the probability distributions (only a shoulder
is visible in $P_{\beta_{\rm max}}(e_{[J]})$), presumably a remnant of the
first-order type transition of the pure system. In the average behaviour, it seems
that at small values of $p$, these types of samples are ``lost'' in the large majority of typical
samples which have a ``second-order type'' of probability distribution.
This observation is corroborated by similar ``signals'' in 
Figs.~\ref{FigHistoChiRare-0.44-128} to \ref{FigHistoChiFoot-0.44-128}
concerning the shape of the susceptibility (narrow peak 
for rare events with large susceptibilities and broader for others),
of the order parameter (sharp increase with $\beta$ at the transition for
rare events, and smoother variation for the typical samples),
or of the Binder cumulant (deep well at the transition in the case of
rare events and less pronounced wells for typical ones).

\begin{figure*}
        \epsfxsize=9.5cm
        \begin{center}
        \mbox{\epsfbox{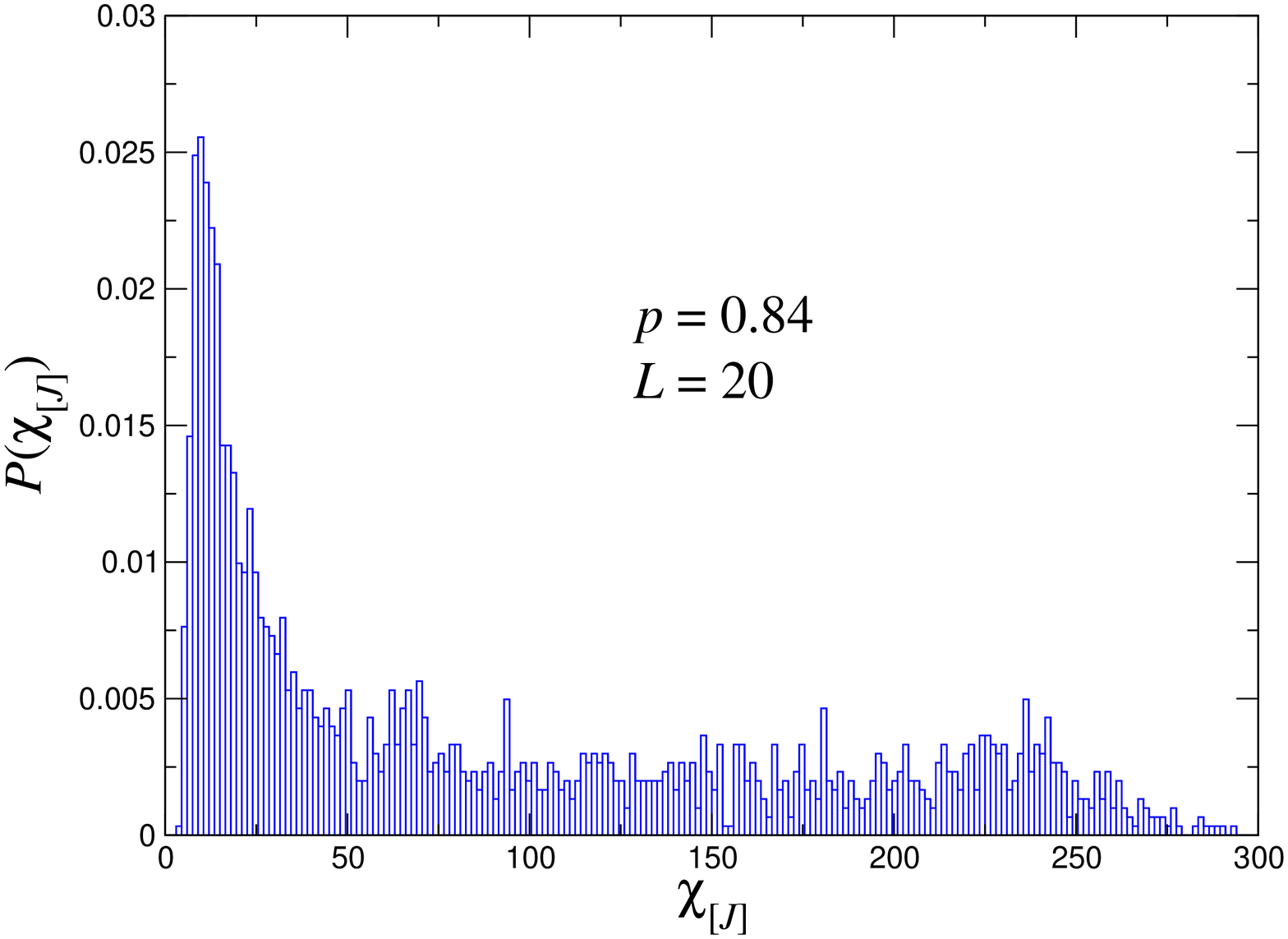}}
        \end{center}\vskip 0cm
        \caption{\small Probability distribution of the susceptibility for a 
         system of size $L=20$ at $p=0.84$ and $\beta J=0.74704$, in the seemingly
 	first-order regime. The simulation was performed with the multibondic algorithm.}
        \label{FigHistoChiL20p084}
\end{figure*}

\begin{figure*}
        \epsfxsize=9.5cm
        \begin{center}
        \mbox{\epsfbox{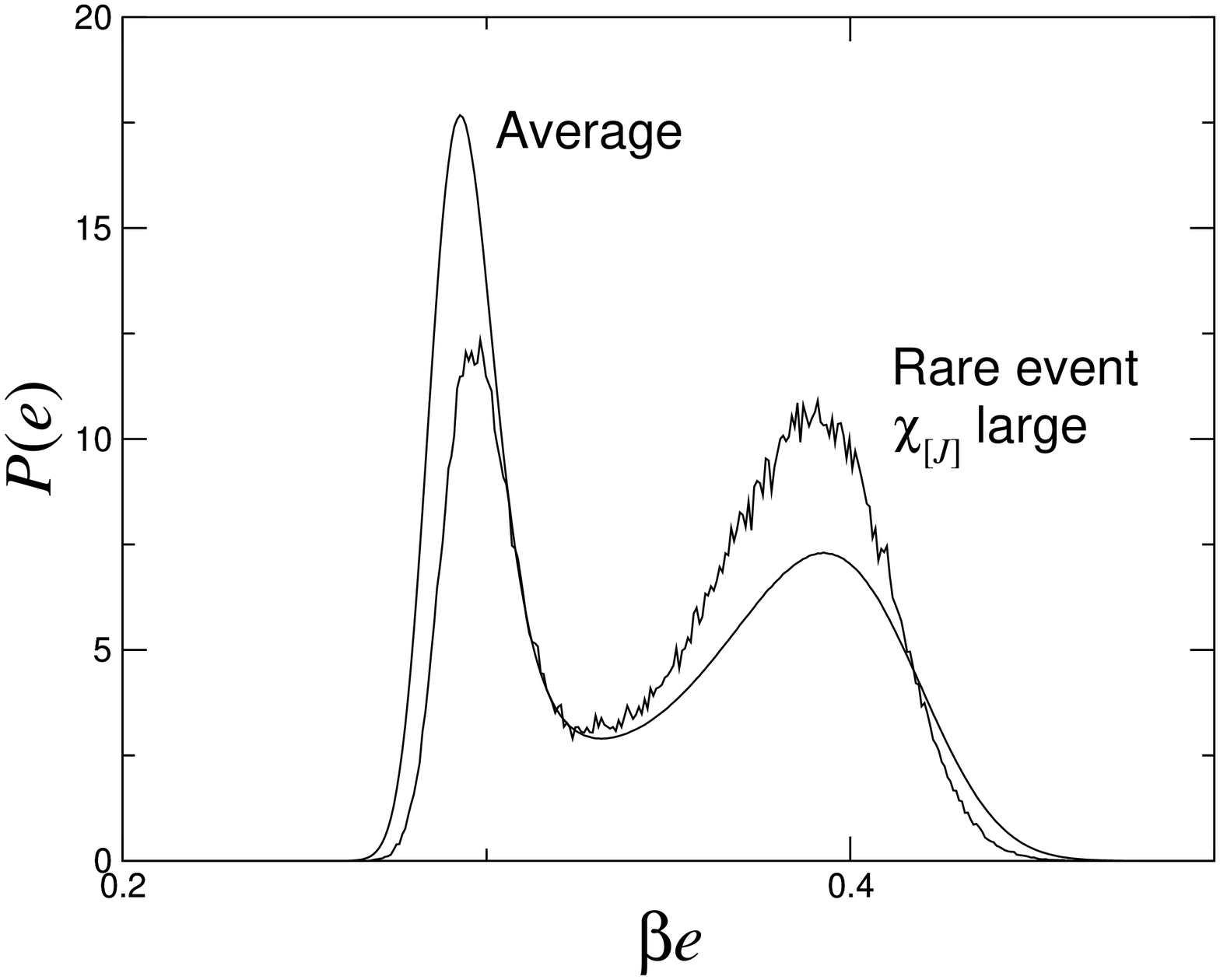}}
        \end{center}\vskip 0cm
        \caption{\small Probability distributions $\overline P$ and $P_{[J]}$
        of the energy $e$ for the average behaviour and for
        a rare event (large susceptibility), respectively, 
        at $p=0.84$ ($L=13$).
        The double-peak structure
        suggests a behaviour for this specific sample which is similar to
        the one observed at a first-order transition.
        The simulation is performed at inverse temperature 
        $\beta J=0.746\,356$.}
        \label{FigEvRarePdeE}
\end{figure*}

We may thus argue that a possible mechanism which keeps the pure model's first-order
character of the transition at larger values of $p$ is connected to
the occurrence of a larger  proportion of samples with the ``first-order type'',
i.e., a very big susceptibility signal at $\beta_{\rm max}$. 
In Fig.~\ref{FigHistoChiL20p084}, the quite long
tail of large susceptibilities in the susceptibility distribution 
confirms this assumption, for  $p=0.84$  ($L=20$), 
i.e., closer to, or probably inside the first-order regime. Also the double-peak
structure of the energy distribution at this dilution
(see Fig.~\ref{FigEvRarePdeE}) is compatible with
a first-order like transition for the average behaviour (of course one
would have to study the evolution of the energy barrier as the size 
increases, but this makes no sense for a specific disorder realization for
which the notion of thermodynamic limit is meaningless).
The possible interpretation is that the rare events of higher susceptibilities
when $p$ becomes larger are more comparable to a system displaying a 
first-order transition. This would explain that the susceptibility peak is 
narrower (and thus does coincide with the temperature of the maximum
of the average only in very rare cases).

\section{Phase diagram and strength of the transition\label{sec5}}
\subsection{Transition line}

We can now come back to the preliminary phases of this work.
The transition temperature was determined for 19 values of the bond
concentration ranging from $p=0.28$ to $p=1.00$ (pure system). We defined an
effective inverse transition temperature $\beta_c(L,p)$ 
at a given lattice size $L$
as the location of the maximum of the average magnetic susceptibility
$\overline\chi$ (see Fig.~\ref{FigTousLesChi}). Any diverging quantity could equally have been chosen but it
turned out that the specific heat was displaying larger statistical errors than
the magnetic susceptibility. Moreover, the stability of the random fixed point
implies a slowly varying specific heat with a critical exponent 
$\alpha\le 0$.\footnote{We expect a stable randomness fixed point at
        large enough dilutions, where the exponent $\alpha$ should be 
        negative hence the singular contribution 
        to the specific heat would not be 
        diverging.}

For each $p$ and $L$, several Monte Carlo simulations were necessary to
get a reasonable estimate of $\beta_c(L,p)$. 
As mentioned before, histogram reweighting was used to
refine the determination. The procedure was applied up to lattice sizes
$L=16$. The resulting phase diagram 
for two different lattice sizes is plotted in
Fig.~\ref{Fig1}. The data appear to be in a remarkable accordance.

The numerical data presented in Fig.~\ref{Fig1} are furthermore
in agreement with the
mean-field prediction $T_c(p)=p T_c(p=1)$ for large bond concentration,
close to the pure system
($p\simeq 1$). At smaller concentration $p$, the topological properties of
the bond configuration become important and the mean-field prediction fails
to reproduce the observed behaviour. The effective-medium approximation
introduced in this context in the eighties
\bfig[h]
        \epsfxsize=8.5cm
        \begin{center}
        \mbox{\epsfbox{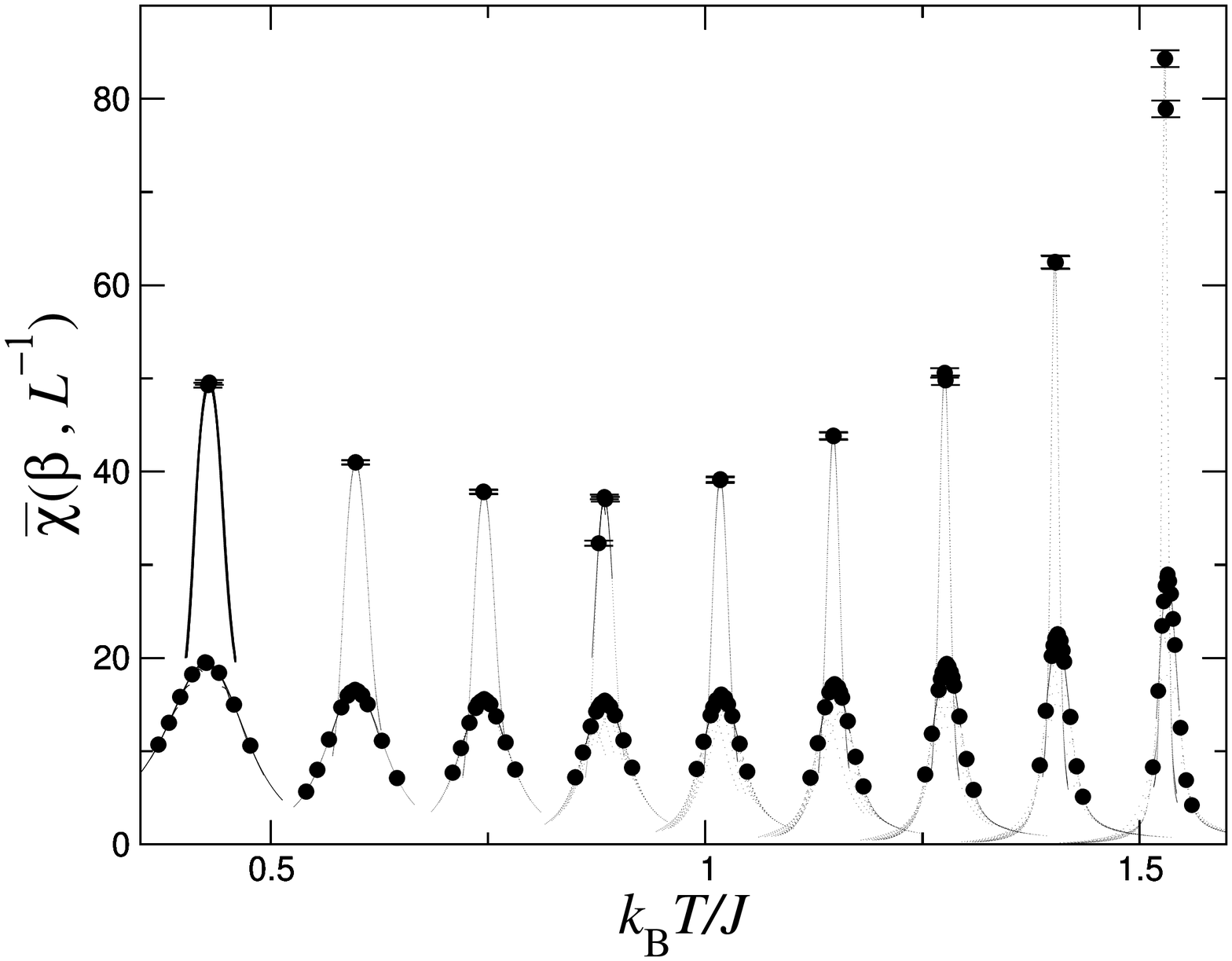}}
        \end{center}\vskip 0cm
        \caption{\small Average susceptibility and its histogram reweighting
        for systems of sizes $10^3$ and
        $16^3$ for dilutions (from left to right) $p=0.32$, 0.40, 0.48, 0.56,
        0.64, 0.72, 0.80, 0.88, and 0.96.}
        \label{FigTousLesChi}
\efig
by Turban~\cite{Turban80} reproduces quite accurately the
numerical data. Limiting the approximation to a single bond, the following
estimate for the transition temperature is obtained:
\begin{equation}
  \beta_c(p)=J^{-1}\ln\left[{(1-p_c)e^{\beta_c^{\rm pure}J}-(1-p)\over 
        (p-p_c)}\right],
  \label{eq7}
\end{equation}
where $\beta_c^{\rm pure}J=0.62863(2)$ for the pure system.
This expression is exact (as exact as it might be with numerical factors
introduced) in the limits of the pure system ($p=1$) 
and the percolation threshold ($p_c = 0.248\,812\,6(5)$).

\bfig
        \epsfxsize=9.5cm
        \begin{center}
        \mbox{\epsfbox{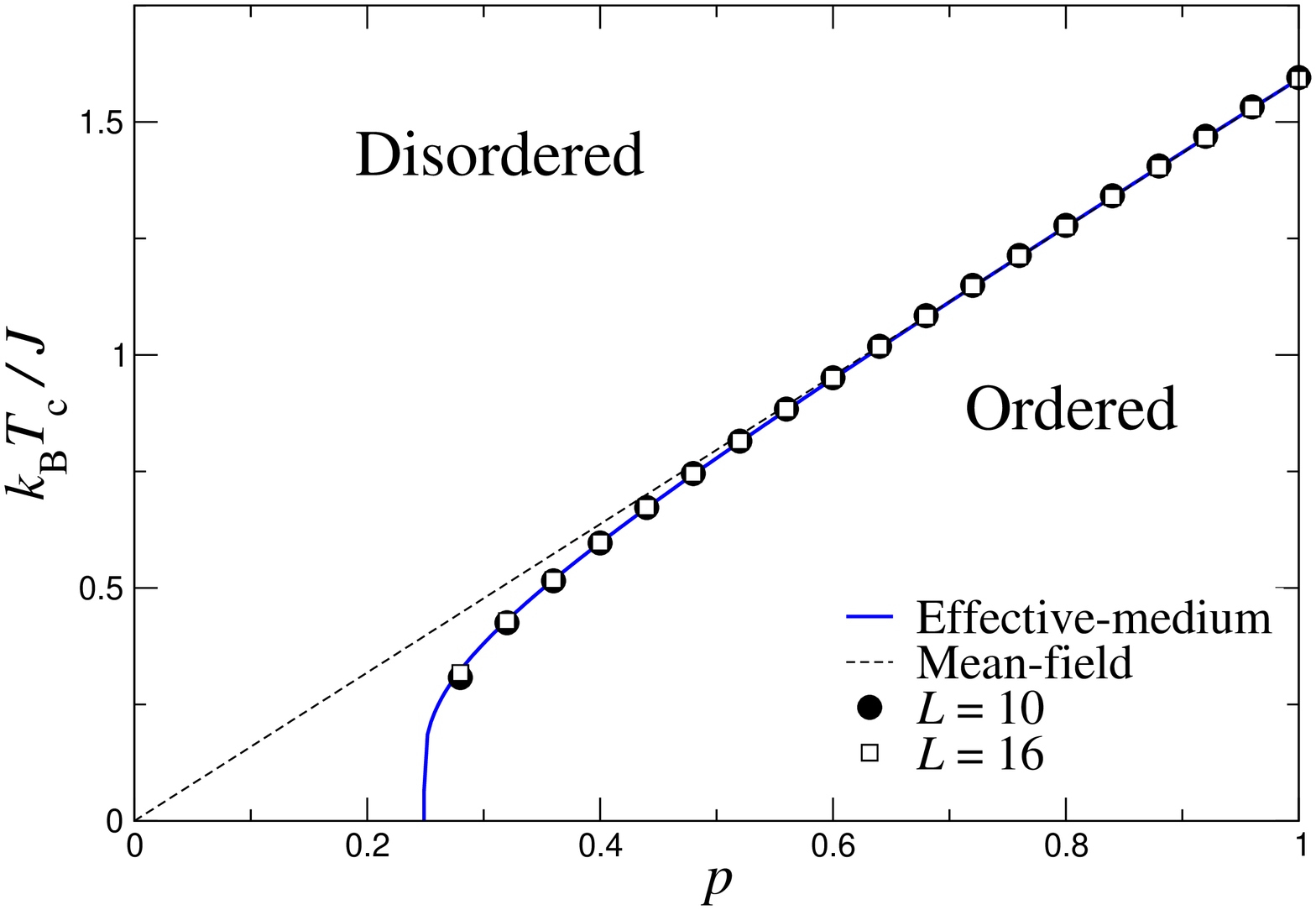}}
        \end{center}\vskip 0cm
        \caption{\small Transition temperatures 
        $k_BT_c(p)/J$ with respect to the
        bond concentration $p$ for two lattice sizes $L=10$ and $L=16$.
        Mean-field and effective-medium approximations are also indicated
        by the dashed and solid lines, respectively.}
        \label{Fig1}
\efig

\subsection{Order of the transition}
Distinguishing a weak first-order phase transition from a continuous one
is a very difficult task. The autocorrelation time of the energy $\tau^e$
at the transition temperature may be useful, since it displays a 
behaviour which depends on the order of the transition. When using a canonical
Monte Carlo simulation for the study of a first-order transition, the
time-scale of the dynamics is dominated by the tunnelling events between
the ordered and disordered phases in coexistence at the transition
temperature. Such a tunnelling event implies the creation and the growth of
an interface whose energy cost behaves as 
$\beta\Delta F = 2\sigma_{\rm o.d.} L^{D-1}$ where $\sigma_{\rm o.d.}$ 
is the reduced interface tension. As a consequence, the autocorrelation
time grows exponentially as
\begin{equation}
  \tau^e(L) \sim e^{{2\sigma_{\rm o.d.}}L^{D-1}}.
  \label{eq8}
\end{equation}
For a continuous phase transition, this interface tension vanishes and
the autocorrelation time scales as a power-law of the lattice size,
\begin{equation}
  \tau^e(L)\sim L^z,
  \label{eq9}
\end{equation}
where $z$ is the dynamical critical exponent.

\bfig
        \epsfxsize=9.5cm
        \begin{center}
        \mbox{\epsfbox{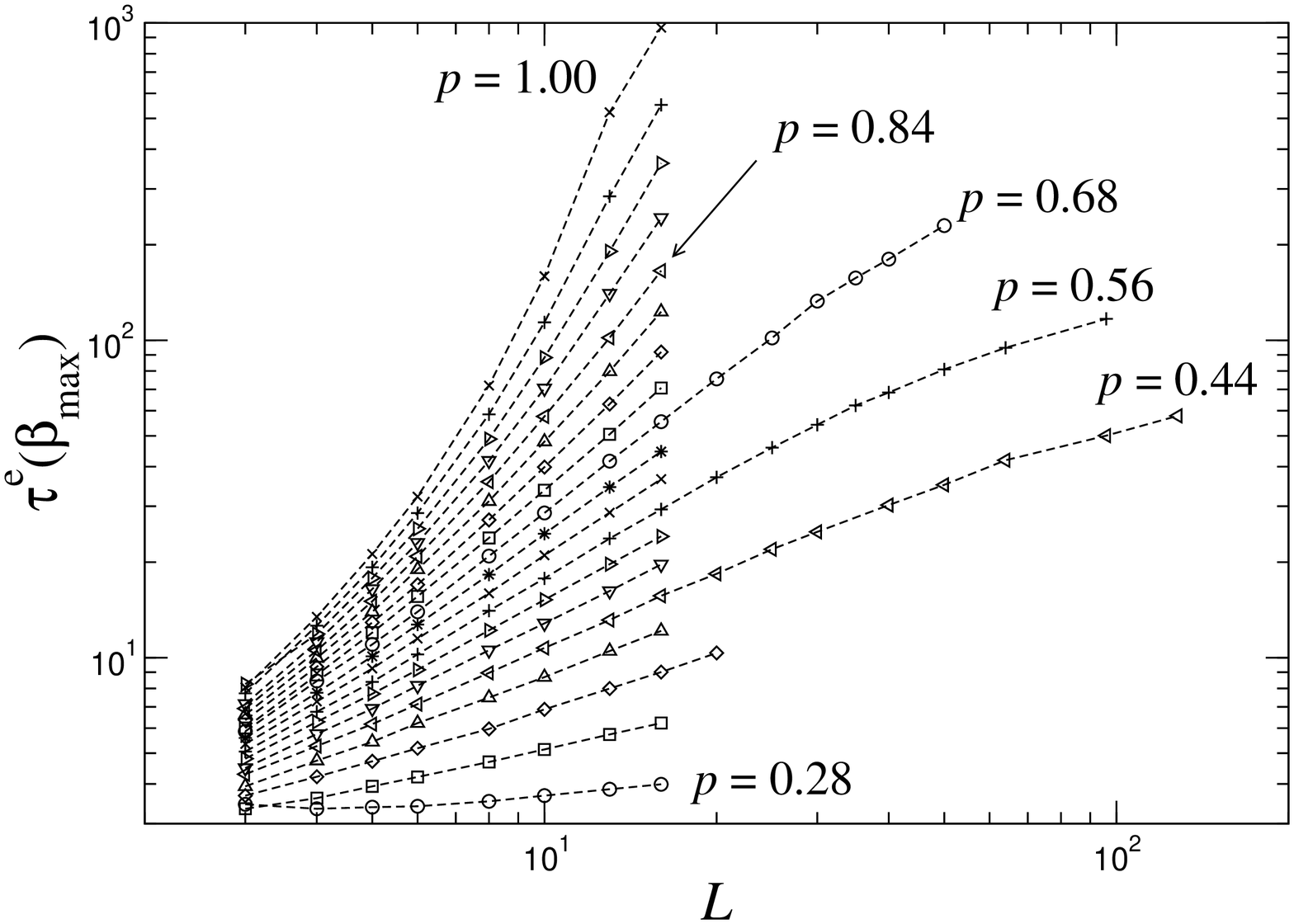}}
        \end{center}\vskip 0cm
        \caption{\small Autocorrelation time of the energy $\tau^e$ with respect
          to the lattice size at the (pseudo-) transition temperature. The curves
          correspond to different bond concentrations $p$ (from bottom
          $p=0.28$ to the top $p=1.00$ in steps of $0.04$). The results 
          shown here all follow
          from MC simulations using the Swendsen-Wang algorithm.}
        \label{Fig3}
\efig

The numerical estimates of the autocorrelation time $\tau^e$ are plotted in
Fig.~\ref{Fig3} for several dilutions. They show a growth of the autocorrelation
time with the lattice size which becomes dramatic as the bond concentration
increases and a behaviour compatible with a power law of the system size when
$p$ decreases, as expected since the dilution softens the transition and thus
reduces the dynamical exponent $z$. Nevertheless, it is not possible to distinguish 
precisely the two regimes on a plot of the autocorrelation time versus the
lattice size. Here, we may locate approximately the boundary between the two regimes
around -- slightly above $p=0.68$. Indeed, the autocorrelation time at $p=0.68$
is very well fitted with a power-law for all lattice sizes smaller than $L=30$.
Above, the data display a downward bending that can be explained by a correction
to the power-law behaviour but not by an exponential prefactor (the bending would be
upward). On the other hand, for $p=0.84$ it is not possible to find any set of
three consecutive points that could be fitted by a power-law: the autocorrelation time
clearly grows faster than a power-law.
Using two successive lattice sizes $L_1$ and $L_2>L_1$, we defined an
effective dynamical exponent
\begin{equation}
  z_{\rm eff}(L_1,L_2)
  ={\ln\tau^e(L_2)-\ln\tau^e(L_1)\over \ln L_2-\ln L_1}
  \label{eq10}
\end{equation}
which is expected to reach a finite value for continuous transitions and to
diverge for first-order ones. The data, plotted in Fig.~\ref{Fig4}, again do
not  lead to any sound estimate of the location of the tricritical point.
Nevertheless, the transition again definitely remains continuous up to the bond
concentration $p=0.68$. For higher concentrations, the data show an increase
of the dynamical exponent with lattice size, but it is not possible
to state unambiguously whether they develop a divergence or not.
We also notice that the necessary finite number of iterations leads to an
underestimate of $\tau^e$ and thus of $z$ for bond concentrations close to $p=1$
at large lattice sizes (this is particularly clear in Fig.~\ref{Fig4} for the
size $L=13-16$). Multi-bondic simulations were thus needed in this case to
improve the measurement of thermodynamic quantities when $p$ is close to 1.

\bfig
        \epsfxsize=8.5cm
        \begin{center}
        \mbox{\epsfbox{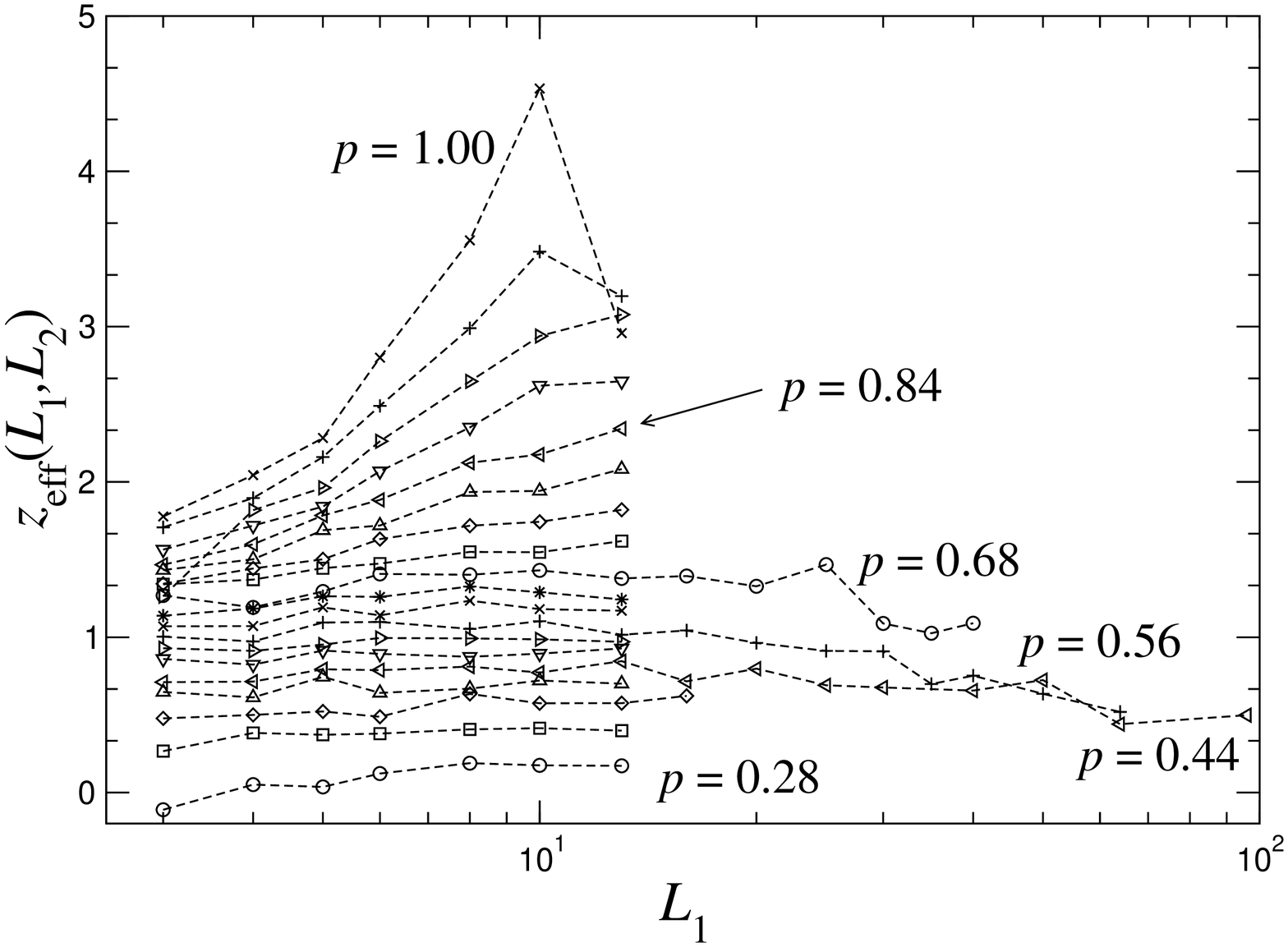}}
        \end{center}\vskip 0cm
        \caption{\small Effective dynamical exponent (SW algorithm) 
        with respect to the smaller lattice size at the transition temperature.
	The curves correspond to different bond concentrations $p$ (from bottom
          $p=0.28$ to the top $p=1.00$ in steps of $0.04$).}
        \label{Fig4}
\efig
\bfig
        \epsfxsize=12.5cm
        \begin{center}
        \mbox{\epsfbox{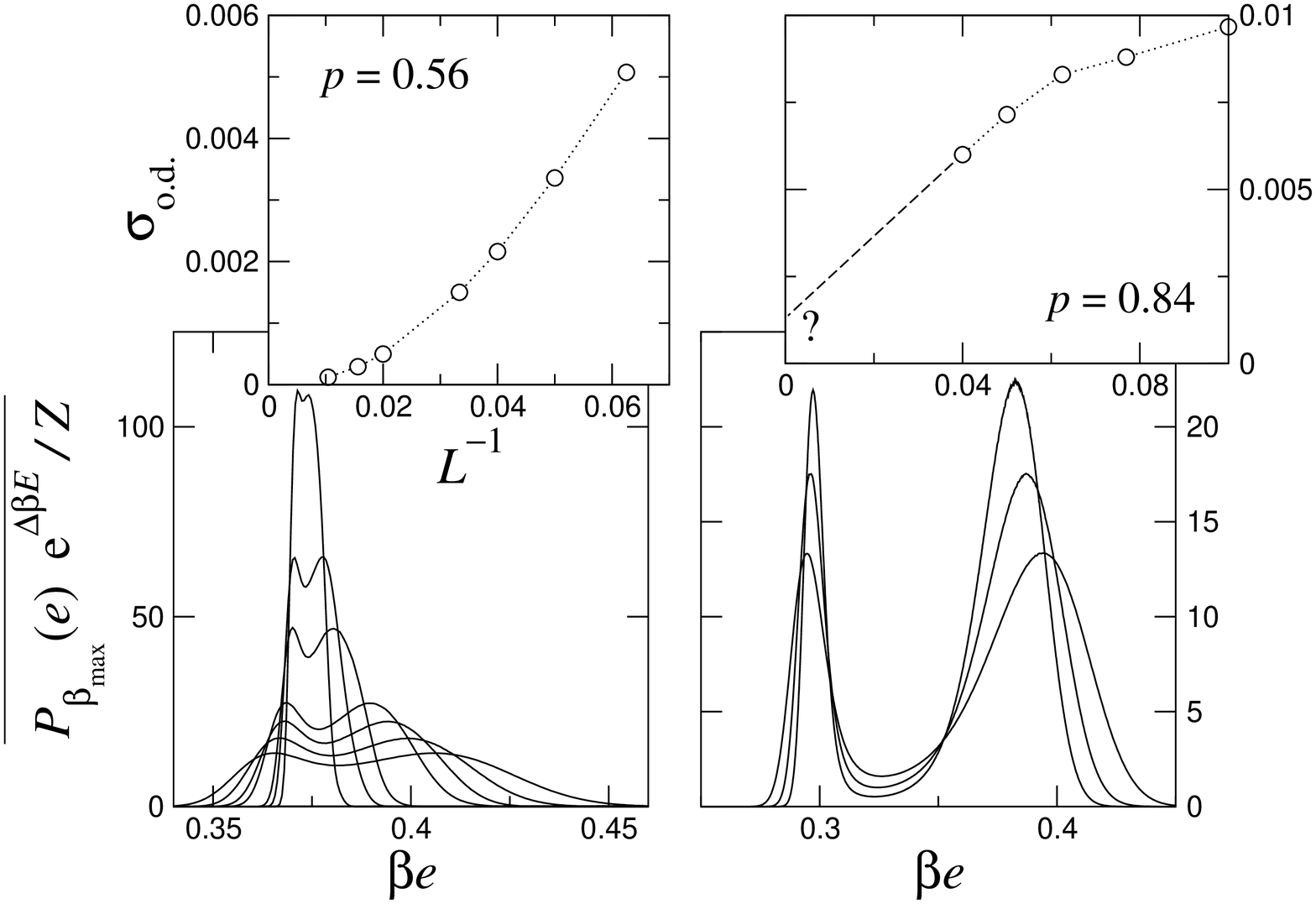}}
        \end{center}\vskip 0cm
        \caption{\small Probability distribution of the energy at the temperature
          for which the two peaks have equal heights. The two plots correspond
          to two different bond concentrations: $p=0.56$ on the left
        (SW algorithm, increasing sizes $L=25$, 30, 35, 40, 50, 64, and 96) and
          $p=0.84$ on the right (multi-bondic simulations, sizes $L=16$, 20, 
        and 25). 
        The order-disorder interface tension
        $\sigma_{\rm o.d.}=\ln (P_{\rm max}/P_{\rm min})/(2L^{D-1})$ is 
        plotted against $L^{-1}$ in the upper part of the figure.}
        \label{fig5}
\efig

Another approach is provided by the behaviour of the order-disorder
interface tension. Numerically, the interface tension can be estimated from the
probability distribution $P(e)$ of the energy. One has
\begin{equation}
  {P_{\rm min} \over P_{\rm max}}
  \propto e^{-{\beta\Delta F}}=e^{-2\sigma_{\rm o.d.} L^{D-1}}.
  \label{eq11}
\end{equation}
Indeed, the free-energy barrier can be related to the ratio of the
(equally high) probabilities of the ordered and disordered 
phases (corresponding to the two peaks) and of the mixed phase regime
involving two interfaces\footnote{Due to the employed periodic boundary
conditions only an even number of interfaces can occur for topological
reasons.} and which corresponds to the bottom of the gap between the two peaks.
We started from the effective transition temperatures estimated from the
maxima of the magnetic susceptibility. At this temperature, the statistical
weight of the ordered and disordered phases are comparable so the height of
the peaks is very different. In order to define the interface tension, we
reweighted the time series of the simulations to the (close) temperature
for which the two peaks have equal heights. The order-disorder interface tension
is plotted against the inverse of the lattice size at the transition temperature
in the upper part of Fig.~\ref{fig5}. It shows undoubtedly a vanishing of the
interface tension for $p=0.56$, and presumably for $p=0.76$ (not shown here)
also, being a clear indication of a disorder induced second-order transition. 
On the other hand, for $p=0.84$ the interface tension seems to converge
towards a finite (but very small?) value in the thermodynamic limit, 
which can be taken as a signal for the persistence of the
first-order nature of the transition in the pure case at $p=1$ 
down to this dilution.

As a consequence, we are led to the conclusion that the tricritical point is 
presumably located between $p=0.68$ and $p=0.84$, the upper bound corresponding
to the observation of an exponential growth of the autocorrelation time and
the lower to a constant dynamical exponent and the vanishing of the latent heat
(both values of $p$ are indicated in the previous Figs.~\ref{Fig3} and \ref{Fig4}).
However, one cannot unambiguously prove by numerical simulations on finite systems that
what we identified as a second-order phase transition is not a weak first-order phase 
transition with a correlation length larger than $L=128$, or that the fast growth of the
autocorrelation time for $p\ge 0.84$ is not a cross-over to a power-law regime at
larger system sizes.

\section{Critical behaviour\label{sec6}}
\subsection{Leading behaviour and critical exponents}
We now concentrate on the second-order regime only, i.e., on
$p\le 0.68$ where we performed an investigation of the universality class
at the disorder fixed point.
The critical exponents are computed using the finite-size 
scaling behaviour of the
physical quantities (Eqs.~(\ref{eq7bisa})-(\ref{eq7bisd}))
at the effective transition temperature $\beta_c(L,p)$. 
In the usual
renormalisation group scheme for disordered systems, the renormalisation flow is
subject to the influence of three fixed points describing respectively the pure
system, the random system and the percolation transition. The scaling 
behaviour is
thus expected to display large corrections resulting in a cross-over to a
unique universal behaviour at large lattice sizes. According to this scheme,
the exponents which are measured are expected to be 
(apparently) concentration dependent. In the previous sections
(see, e.g., Fig.~\ref{FigTousLesChi}), the corrections
to scaling for the transition temperature have been observed to be weaker
for the bond concentration $p=0.56$. This behaviour is illustrated, e.g., in 
Fig.~\ref{FigBeta_c_vs_L} where the cross-over effect
reflects in the bending of the curves $\beta_{\rm max}(L,p)J$ vs. $L^{-1}$ for three
dilutions $p=0.32$, $p=0.56$, and $p=0.80$. 
The curve at $p=0.56$, on the other hand, is almost {\it flat}.
The corresponding data for the
three main dilutions in the second order regime,
$p=0.44$, $p=0.56$, and $p=0.68$, are then
plotted against $L^{-1/\nu}$ on the right part. Although the value of
$\nu$ is not yet known, we anticipate here the later result, using 
already the ``to-be-determined-exponent''. Again, the curve at $p=0.56$
has an almost vanishing slope.
As a consequence, we decided to
make further large-scale Monte Carlo simulations
at this concentration $p=0.56$ 
up to the lattice size $L=96$. To monitor the effects
of the competing fixed points, we also made additional large-scale 
simulations for the
concentrations $p=0.44$ (towards the percolation transition) and $p=0.68$ 
(towards the regime of first-order transitions) up to the lattice sizes $L=128$
and $L=50$, respectively (size limitations  at these concentrations are
linked to the discussion of Sect.~\ref{sec3}).

\bfig
        \epsfxsize=10.5cm
        \begin{center}
        \mbox{\epsfbox{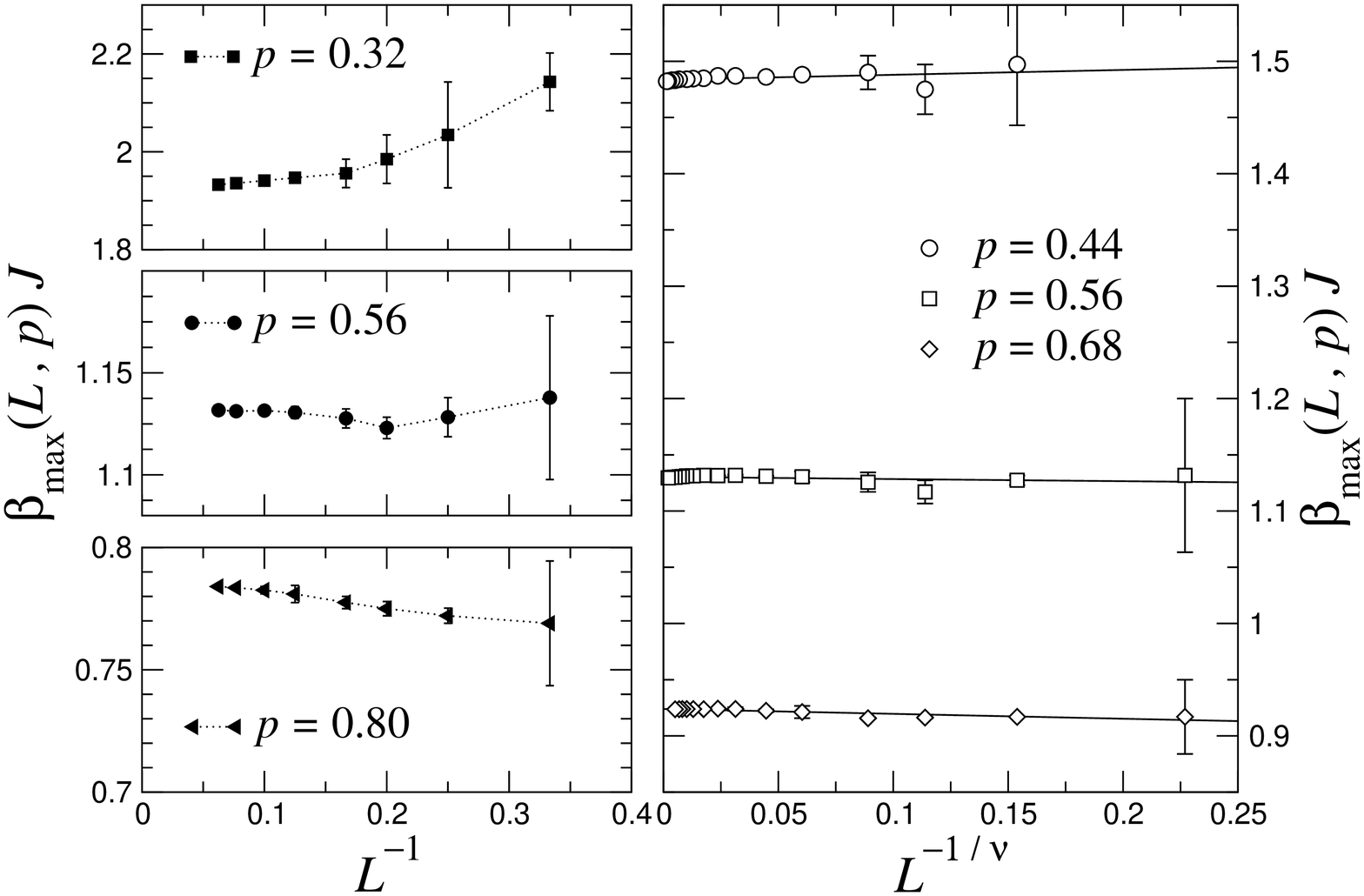}}
        \end{center}\vskip 0cm
        \caption{\small Evolution of the size-dependent (pseudo-)critical coupling
        with the inverse system size for relatively small sizes on the left
        plots. 
        The same on the right plot for the three main dilutions, where 
        the data are by anticipation fitted to a linear relation
        $\beta_{\rm max}(L,p)=\beta_c(p)+aL^{-1/\nu}+\dots$, where our
        estimate for $\nu$ ($\approx 0.75$) will be discussed later. 
        The slope coefficient is 
        slightly positive for $p=0.44$, slightly negative for $p=0.68$ and
        virtually zero at $p=0.56$, where the corrections-to-scaling
        (at least for this quantity) appear to be the smallest.}
        \label{FigBeta_c_vs_L}
\efig

\bfig[ht]
        \epsfxsize=9.cm
        \begin{center}
        \mbox{\epsfbox{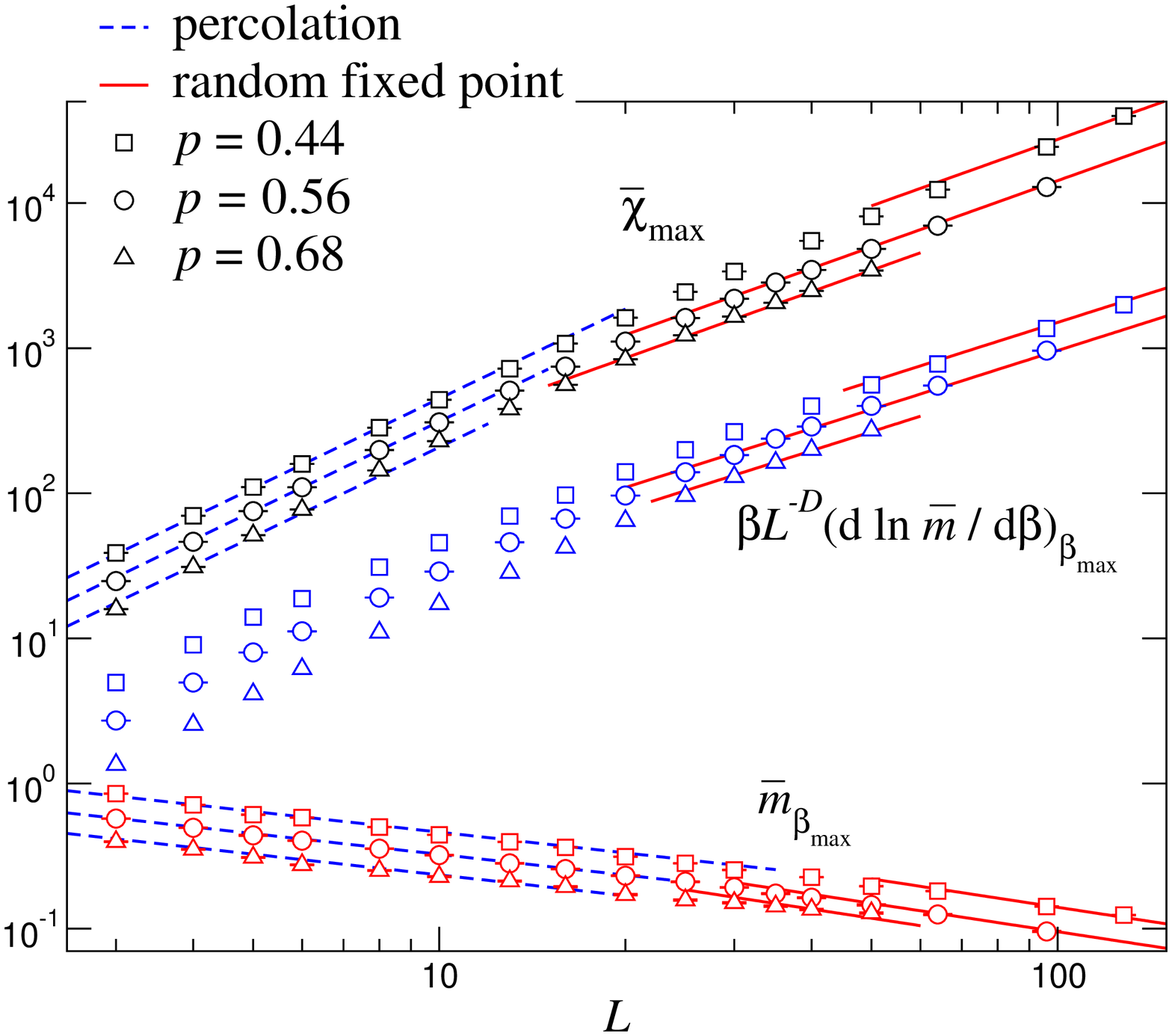}}
        \end{center}\vskip 0cm
        \caption{\small Finite-size scaling behaviour of the 
        susceptibility, the magnetisation and of $\beta L^{-D} d\ln\overline m/d\beta$  
        at $\beta_{\rm max}$ (the quantities
        have been shifted in the vertical direction for the sake of clarity). 
        The behaviour at small lattice sizes is
        presumably governed by the percolation fixed point (shown as 
        dashed lines and characterised by exponent ratios
        $\gamma/\nu\simeq 2.05$ and $\beta/ \nu\simeq 0.475$). Above a
        crossover length scale a new (random) fixed point is reached (shown
        by the 
        solid lines, with exponent ratios $\gamma/\nu\simeq 1.535$,
        $1/\nu\simeq 1.34$,  and $\beta/ \nu\simeq 0.73$, discussed in
        detail below).}
        
        \label{FigChiMax-vs-L}
\efig
\bfig
        \epsfxsize=9.cm
        \begin{center}
        \mbox{\epsfbox{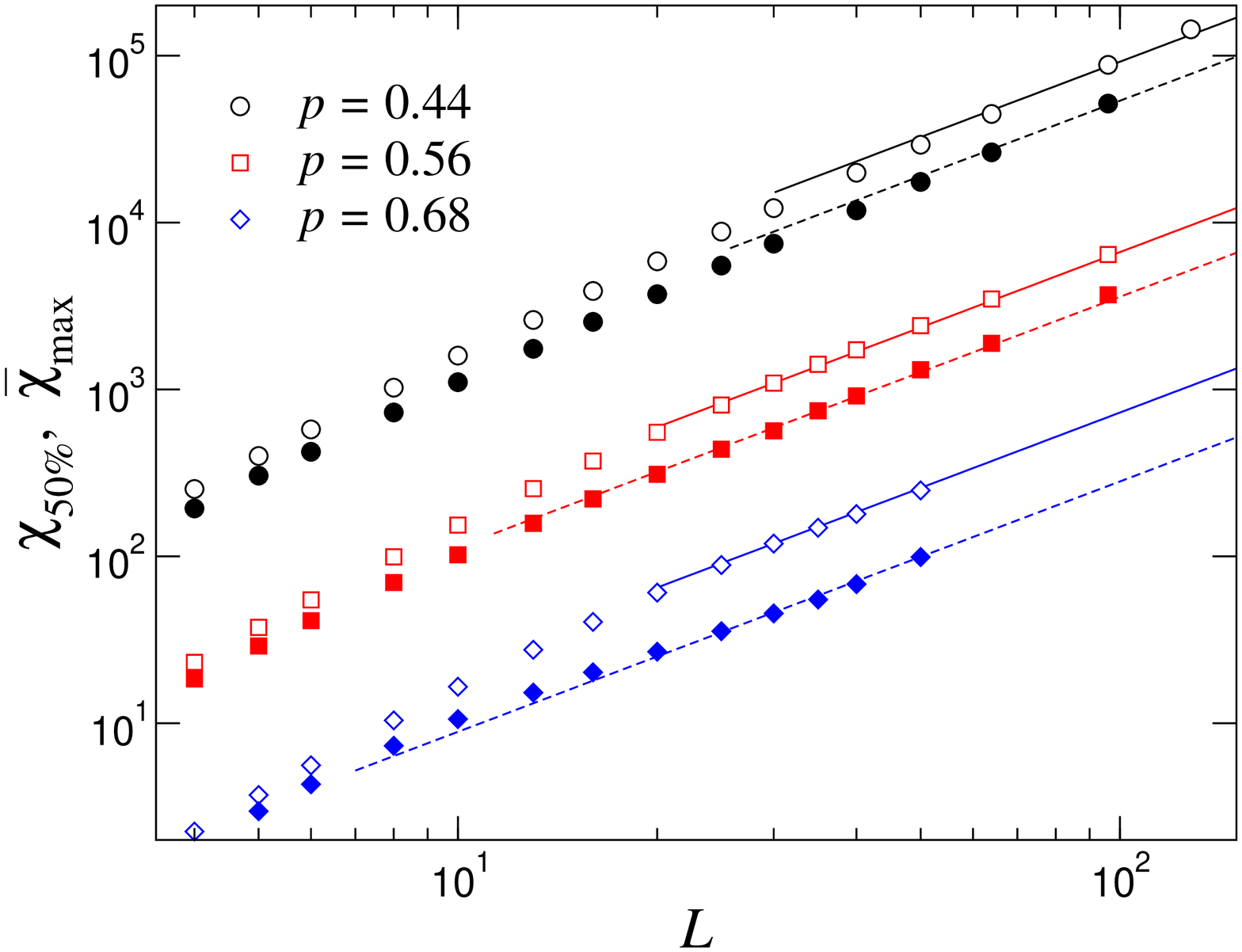}}
        \end{center}\vskip 0cm
        \caption{\small Log-log plot of the average 
        susceptibility (open symbols) and the typical
        susceptibility (filled symbols), 
        as defined in Eq.~(\ref{chi50}) by $\chi_{50\%}$ for the three
        principal dilutions studied, indicating that the asymptotic
        scaling regime sets in earlier for the latter quantity.}
        \label{FigChi3p}
\efig

In Fig.~\ref{FigChiMax-vs-L}, 
the finite-size scaling behaviour of
the maximum susceptibility, $\overline \chi_{\rm max}$, the magnetisation
at $\beta_{\rm max}$ and the derivative of $\ln\overline m$ 
with respect to the inverse temperature evaluated at ${\beta_{\rm max}}$
are plotted versus
the system size on a log-log scale. 
These curves should give access
to the  exponents $\gamma/\nu$, $\beta/\nu$, and $1/\nu$, respectively.
The three main dilutions are represented. One clearly observes a crossover 
between two regimes. 
For small lattice sizes, the system is strongly influenced by the
proximity of a perturbing fixed point while a different, unique fixed
point, is apparently reached at large sizes, as revealed by the slopes
which are at first sight independent of the dilution when the linear extent
of the lattice reaches values of about $L\ge 30$.
The most probable susceptibility $\chi_{50\%}$ is shown in Fig.~\ref{FigChi3p}
and can also lead 
to estimates for $\gamma/\nu$.
According to the discussion given in Sect.~\ref{sec3}, 
we expect that the most probable
susceptibility is better described than the average susceptibility, for
which there exists a significant contribution of rare events, and these
rare disorder realizations might be poorly scanned if a too small number
of samples is considered. This difficulty might be circumvented through the
study of what we defined as $\chi_{50\%}$ in Eq.~(\ref{chi50}). 
In the presence of multifractality, the universal behaviour of 
$\chi_{50\%}$ should differ from that of $\overline \chi$. Since such
a peculiar behaviour does not occur in the case of 
a global quantity~\cite{Derrida84}, like $\chi$, we expect compatible
values of $\gamma/\nu$ as deduced from  $\chi_{50\%}$ or $\overline\chi$.
Observing the data plotted in 
Fig.~\ref{FigChi3p}, in fact, confirms our previous analysis. It seems that
$\chi_{50\%}$ is less influenced by the crossover effects than 
the average $\overline\chi_{\rm max}$. 
In order to support this statement,
we will present the results of fits of the susceptibility in two 
different tables for the two regimes and for the three main dilutions:

\begin{itemize}
\item[$\triangleright$] At small lattice sizes, the behaviour of 
$\overline \chi_{\rm max}$ and $\overline m_{\beta_{\rm max}}$
is in all three cases compatible with the percolation exponents
$(\gamma/\nu)_{\rm perco}\simeq 2.05$ and 
$(\beta/\nu)_{\rm perco}\simeq 0.475$ shown in Fig.~\ref{FigChiMax-vs-L} 
by the dashed lines. This
seems to be true (particularly in the case of the susceptibility) 
over a wider range of sizes for $p=0.44$ than for $p=0.56$ or
$p=0.68$. This observation is compatible with a stronger influence of the
percolation fixed point when $p=0.44$, which is closer to the percolation
threshold than the two other dilutions.
Surprisingly, the assumption of a percolation influence is absolutely
not confirmed\footnote{The
expected percolation exponent would be $1/\nu\simeq 1.124$ while the slope at
small sizes is larger than in the random regime where it takes a value
close to $1/\nu\simeq 1.35$.}
by the behaviour at small sizes of the third quantity of
interest, $L^{-D}(d\ln\overline m/d\beta)_{\beta_{\rm max}}$.
Due to the involved differentiation with respect to inverse temperature,
the identification with percolation quantities becomes less obvious, but
we do not have any
explanation for this strange result.
In Table~\ref{TabApp4}, we try to  point out
the {\em influence of the percolation fixed point}. This is achieved by 
power-law fits between a fixed minimum size $L_{\rm min}=4$ up to an 
increasing maximum size $L_{\rm max}$ below the value $L=30$ which
apparently marks the modification in the behaviour of the physical quantities
under interest.
We first observe that $\beta/\nu$ starts
from a value very close to the percolation value, and second, that 
$\chi_{50\%}$ has always a lower exponent (i.e. more distinct from the
percolation value).

\begin{center}
\begin{table}[h]
        \caption{\small Exponents deduced from the finite-size scaling behaviour of
         $\overline\chi_{\rm max}$ and  $\chi_{50\%}$ in the vicinity of the 
        percolation fixed point (small sizes). Recall the percolation
        value $(\gamma/\nu)_{\rm perco}\simeq 2.05$ for comparison.
        \label{TabApp4}}\small
        \bcenter\begin{tabular}{llllllll}
        \noalign{\vspace{3mm}}
        \hline\noalign{\vspace{0.4pt}}
        \hline\noalign{\vspace{0.4pt}}
        \hline
              &               & \multicolumn{2}{c}{$p=0.44$}             &
                               \multicolumn{2}{c}{$p=0.56$}              &
                               \multicolumn{2}{c}{$p=0.68$}              \\
              &               & \multicolumn{2}{c}{$\gamma/\nu$ deduced from}
              &
                               \multicolumn{2}{c}{$\gamma/\nu$ deduced from}
              &
                               \multicolumn{2}{c}{$\gamma/\nu$ deduced from}
              \\
              &               & \crule{2} & \crule{2} & \crule{2} \\
$L_{\rm min}$ & $L_{\rm max}$ &\tvi 06 $\overline\chi_{\rm max}$ & 
        $\chi_{50\%}$ 
                              & $\overline\chi_{\rm max}$ & $\chi_{50\%}$ 
                              & $\overline\chi_{\rm max}$ & $\chi_{50\%}$ \\
\hline
4 & 8 & 2.015 & 1.902 & 2.098 & 1.916 & 2.211 & 1.915 \\
4 & 13& 1.984 & 1.866 & 2.034 & 1.818 & 2.132 & 1.720 \\
4 & 20& 1.954 & 1.833 & 1.973 & 1.748 & 2.051 & 1.579 \\
4 & 30& 1.924 & 1.808 & 1.913 & 1.691 & 1.974 & 1.500 \\ 
        \hline\noalign{\vspace{0.4pt}}
        \hline\noalign{\vspace{0.4pt}}
        \hline
\end{tabular}\\[0.5cm]
\ecenter\end{table}
\end{center}

\begin{center}
\begin{table}[h]
        \caption{\small Exponents deduced from the finite-size scaling behaviour of
         $\overline\chi_{\rm max}$ and  $\chi_{50\%}$ in the vicinity of the 
        random fixed point (large sizes). The largest size taken into
        account in the fits is $L_{\rm max}=128$ for $p=0.44$,
        96 for $p=0.56$, and 50 for $p=0.68$.
        \label{TabApp5}}\small
        \bcenter\begin{tabular}{lllllll}
        \noalign{\vspace{3mm}}
        \hline\noalign{\vspace{0.4pt}}
        \hline\noalign{\vspace{0.4pt}}
        \hline
                             & \multicolumn{2}{c}{$p=0.44$}             &
                               \multicolumn{2}{c}{$p=0.56$}              &
                               \multicolumn{2}{c}{$p=0.68$}              \\
                             & \multicolumn{2}{c}{$\gamma/\nu$ deduced from}
              &
                               \multicolumn{2}{c}{$\gamma/\nu$ deduced from}
              &
                               \multicolumn{2}{c}{$\gamma/\nu$ deduced from}
              \\
        & \crule 2 & \crule 2 & \crule 2  \\
$L_{\rm min}$  & \tvi 06 $\overline\chi_{\rm max}$ & $\chi_{50\%}$ 
                              & $\overline\chi_{\rm max}$ & $\chi_{50\%}$ 
                              & $\overline\chi_{\rm max}$ & $\chi_{50\%}$ \\
\hline
20 &  1.724 & 1.672 & 1.571 & 1.579 & 1.541 & 1.412 \\
25 &  1.711 & 1.664 & 1.543 & 1.587 & 1.479 & 1.471 \\ 
30 &  1.706 & 1.669 & 1.518 & 1.596 & 1.438 & 1.539 \\
35 &   -    &  -    & 1.500 & 1.581 & 1.447 & 1.645 \\
40 &  1.703 & 1.679 & 1.502 & 1.587 & 1.464 & 1.675 \\
50 &  1.695 & 1.657 & 1.506 & 1.593 & \\
64 &  1.680 & 1.659 &       &       & \\
        \hline\noalign{\vspace{0.4pt}}
        \hline\noalign{\vspace{0.4pt}}
        \hline
\end{tabular}\\[0.5cm]
\ecenter\end{table}
\end{center}

\item[$\triangleright$] At large sizes, for each quantity considered here, the curves
corresponding to the three dilutions  in Figs.~\ref{FigChiMax-vs-L} 
and \ref{FigChi3p}
evolve, after a crossover regime
whose exact location depends on the value of $p$, towards a
presumably unique
power-law behaviour which seems to remain stable 
then (solid lines in Fig.~\ref{FigChiMax-vs-L}). We thus believe
that we have reached large enough sizes in order to get  reliable estimates
of the {\em random fixed point exponents}. This is only a visual impression,
since in fact the effective exponents are still subject to significant
variations, especially for the extreme dilutions $p=0.44$ and $p=0.68$.
Effective exponents $\gamma/\nu$, $\beta/\nu$, and $1/\nu$ may be defined
from power-law fits of $\overline \chi_{\rm max}$, $\overline m_{\beta_{\rm max}}$,
and $L^{-D}d\ln\overline m/d\beta$ between an increasing 
minimum size, $L_{\rm min}$, and a maximum one, $L_{\rm max}$.
The value $L_{\max}$ is kept to the maximum available value $L=128$,
$96$, and $50$ for $p=0.44$, 0.56, and 0.68, respectively, and the results
for the susceptibility are presented in Table~\ref{TabApp5}. We see there that 
$\chi_{50\%}$ is again better behaved (more stable) 
than the average susceptibility.

\end{itemize}

Since we are mainly interested in the randomness fixed point, we now
concentrate on fits at large system sizes. An exhaustive summary 
(i.e. for all three dilutions under interest) of the 
results of the fits
performed at dilutions $p=0.44$, $p=0.56$, and $p=0.68$ is presented 
in Table~\ref{TabApp3}. The corresponding 
effective exponents are also 
plotted against $L^{-1}_{\rm min}$ in Fig.~\ref{FigExpstEff2}.
These results show that the data analysis is much more complicated than our
previous preliminary determination of exponents in Table~\ref{TabApp5}.
Again, the crossover between percolation and random fixed point
behaviours is visible through the variation of effective exponents and
the data present large corrections-to-scaling. 

\begin{center}
\begin{table}[t]
        \caption{\small Linear fits for $\overline\chi_{\rm max}$, 
        $\overline m_{\beta_{\rm max}}$, and $L^{-D}d\ln\overline m/d\beta$ 
        at $\beta_{\rm max}$, leading to finite-size estimates of the
        combinations of critical exponents $\gamma/\nu$, $\beta/\nu$
        and $1/\nu$. 
        These results correspond to the three main dilutions, 
        and they are extracted from
        the finite-size scaling behaviour of the quantities at the
        temperature where the maximum of the average 
        susceptibility is found by histogram reweighting. The results for
        dilutions $p=0.44$ and $p=0.68$ are less stable than for $p=0.56$,
        reflecting the role of the crossover.
        \label{TabApp3}}
        \small
        \bcenter\begin{tabular}{llllllllll}
        \noalign{\vspace{3mm}}
        \hline\noalign{\vspace{0.4pt}}
        \hline\noalign{\vspace{0.4pt}}
        \hline
        $p$ & $L_{\rm min}$ & $L_{\rm max}$ & $\gamma/\nu$ & error &
        $\beta/\nu$ & error & $1/\nu$ & error & $\gamma/\nu+2\beta/\nu$\\
        \hline
0.44 & 30 & 128 & 1.706 & 0.006 & 0.544 & 0.005 & 1.395 & 0.006 & 2.794(16) \\
---  & 40 & --- & 1.703 & 0.008 & 0.552 & 0.007 & 1.381 & 0.008 & 2.807(22) \\
---  & 50 & --- & 1.695 & 0.010 & 0.540 & 0.009 & 1.358 & 0.010 & 2.775(28) \\
---  & 64 & --- & 1.680 & 0.016 & 0.534 & 0.014 & 1.357 & 0.016 & 2.748(44) \\
        \hline
0.56 & 30 &  96 & 1.518 & 0.011 & 0.588 & 0.010 & 1.389 & 0.011 & 2.694(31) \\
---  & 35 & --- & 1.500 & 0.014 & 0.592 & 0.012 & 1.362 & 0.013 & 2.684(38) \\
---  & 40 & --- & 1.502 & 0.016 & 0.608 & 0.015 & 1.353 & 0.016 & 2.718(46) \\
---  & 50 & --- & 1.506 & 0.026 & 0.645 & 0.024 & 1.330 & 0.025 & 2.796(74) \\
        \hline
0.68 & 25 &  64 & 1.479 & 0.021 & 0.343 & 0.015 & 1.505 & 0.021 & 2.165(51) \\
---  & 30 & --- & 1.438 & 0.031 & 0.344 & 0.022 & 1.453 & 0.030 & 2.126(75) \\
---  & 35 & --- & 1.447 & 0.047 & 0.342 & 0.033 & 1.437 & 0.046 & 2.13(11)  \\
---  & 40 & --- & 1.464 & 0.075 & 0.547 & 0.051 & 1.379 & 0.075 & 2.56(18)  \\
        \hline\noalign{\vspace{0.4pt}}
        \hline\noalign{\vspace{0.4pt}}
        \hline
\end{tabular}\\[0.5cm]
\ecenter\end{table}
\end{center}

\bfig[h]
        \epsfxsize=9.cm
        \begin{center}
        \mbox{\epsfbox{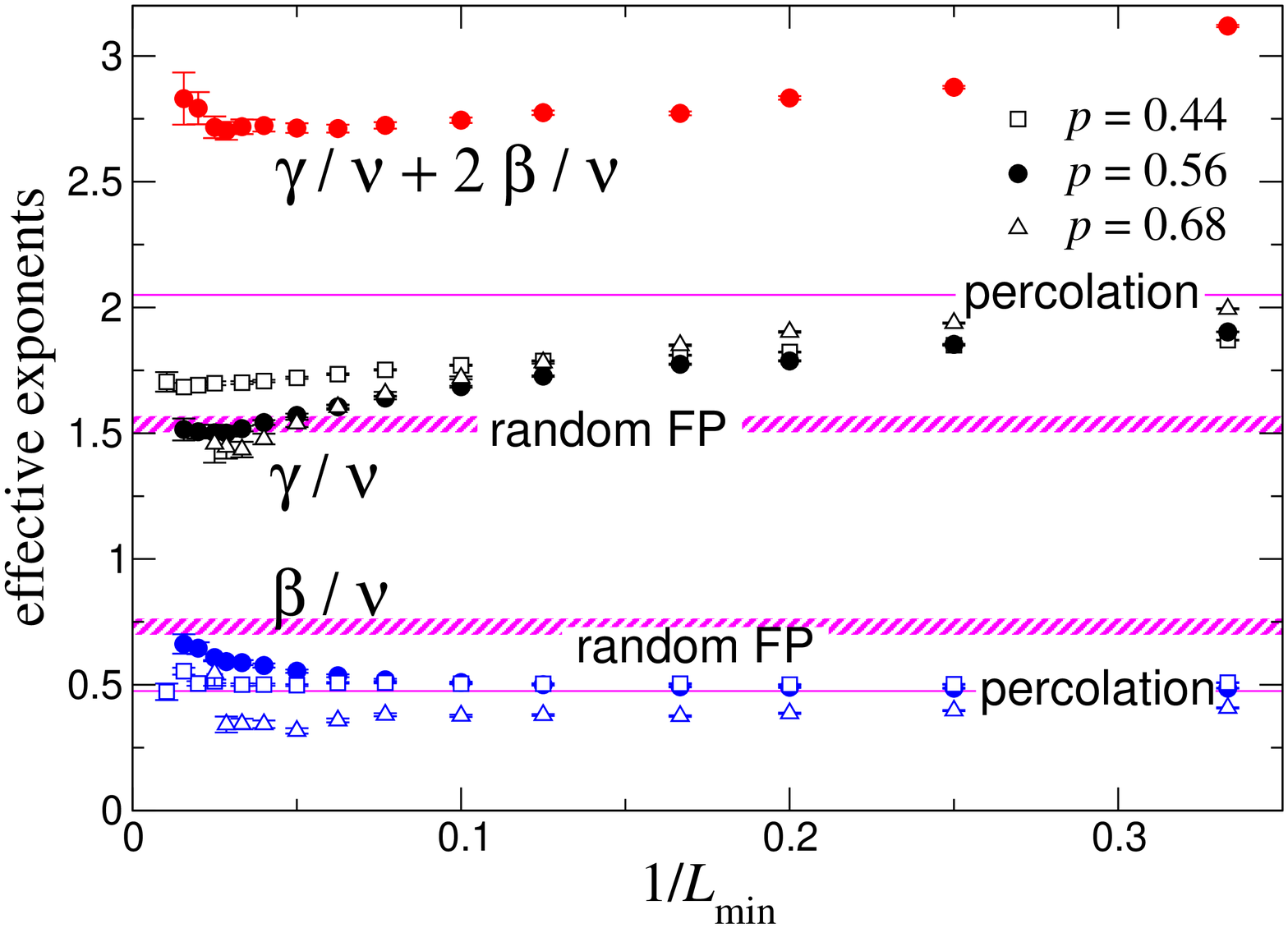}}
        \end{center}\vskip 0cm
        \caption{\small Effective critical 
        exponents $\gamma/\nu$ and $\beta/\nu$, as
        computed from a power-law fit between $L_{\rm min}$ and $L_{\rm max}$,
        with $L_{\max}$ fixed to the maximum available value $L=128$,
        $96$ and $50$ for $p=0.44$, 0.56, and 0.68, respectively. They are
        plotted against $L^{-1}_{\rm min}$. The thin solid line shows the
        percolation values and the shadow stripe corresponds to
        our estimate for the random fixed point values. In the case of the
        dilution $p=0.56$, the value of $2\beta/\nu +\gamma/\nu$ is also
        shown.}
        \label{FigExpstEff2}
\efig

A precise determination of the magnetic exponents is quite difficult. Indeed,
as can be seen in Fig.~\ref{FigExpstEff2}, the effective critical
exponents $(\gamma/\nu)_{\rm eff}$
and $(\beta/\nu)_{\rm eff}$ do not converge towards 
$p$-independent limits when $L_{\rm min} \rightarrow L_{\rm max}$. 
The cross-over effects on the thermal quantities are much smaller. Indeed,
the effective critical exponent $\nu_{\rm eff}$ is converging to a roughly
$p$-independent limit when $L_{\rm min} \rightarrow L_{\rm max}$.
We can give the following estimates for $\gamma/\nu$ and $1/\nu$~:
\begin{eqnarray}
p=0.44: & (\gamma/\nu)_{\rm eff} \simeq 1.68(2), & (1/\nu)_{\rm eff}
\simeq 1.36(2),\\
p=0.56: & (\gamma/\nu)_{\rm eff} \simeq 1.51(3), & (1/\nu)_{\rm eff}
\simeq 1.33(3),\\
p=0.68: & (\gamma/\nu)_{\rm eff} \simeq 1.46(8), & (1/\nu)_{\rm eff}
\simeq 1.38(8),
\end{eqnarray}
simply corresponding to the last line of Table~\ref{TabApp3}, i.e., to the
largest studied value of $L_{\rm min}$, for each dilution.
The value of $\beta/\nu$ on the other hand is definitely not stable and
more subject to the competing influence of fixed points.
For $p=0.44$ for example, the estimate of $(\beta/\nu)_{\rm eff}$  is relatively
stable against variations of $L_{\rm min}$, with fitted values below $0.5$, 
close to the expected value for the percolation transition ($0.475$). This is a quantitative
indication that the system is probably still
subject to cross-over caused by the percolation fixed point.
In the case of $p=0.68$, the estimate of $(\beta/\nu)_{\rm eff}$
is very small, then suddenly increasing for $L_{\rm min}=64$. 
These remarks are consistent with the renormalisation scheme described above. 
In order to help us to decide between the different effective 
values measured at the three dilutions, we use the scaling relation
$\gamma/\nu + 2\beta/\nu = D = 3$ which is almost satisfied  for the bond
concentration $p=0.56$ only (shown in Fig.~\ref{FigExpstEff2})
when taking into account the lattice sizes
$L\ge 50$. For the bond concentrations $p=0.44$ and $p=0.68$, this
scaling relation is not satisfied for any of the accessible
values. One is thus led to conclude that the critical regime has
not yet been reached for these concentrations, in spite of our efforts
to go up to very large sizes.
Remember also that the corrections-to-scaling were found to be the smallest
at $p=0.56$, so the asymptotic regime in neighbouring dilutions should be 
more difficult to reach.
Figure \ref{FigExpstEff2} thus suggests to rely only on the values measured at
dilution $p=0.56$, {\em extrapolated to $L_{\rm min}\to\infty$}, as shown in 
Fig.~\ref{FigExpstEffp0.56}, where a dashed stripe emphasises such an
extrapolation of the effective exponents measured at the largest sizes. 
The values of $(\gamma/\nu)_{\rm eff}$ and 
$(1/\nu)_{\rm eff}$ are indeed stable in the regime $L\ge 35$. We may
thus have {\em reliable estimates} of the asymptotic values for these 
exponents, and a {\em reasonable estimate} for $\beta/\nu$, ratifying the
scaling relation. 

Using this extrapolation procedure, our final estimates of the critical exponents
of the disorder induced random fixed point of the three-dimensional bond-diluted
4-state Potts model are the following values: 
\begin{eqnarray}
& \gamma/\nu & =1.535(30),  \\
& \beta/\nu  & =0.732(24), \\
& 1/\nu      & =1.339(25),
\end{eqnarray}
resulting from a linear extrapolation of the data points for $L_{\rm min} = 25$,
30, 35, 40, 50, and 64 at $p=0.56$. Note that since the data are correlated, we have
kept the error of the last point.

\bfig[t]
        \epsfxsize=9.cm
        \begin{center}
        \mbox{\epsfbox{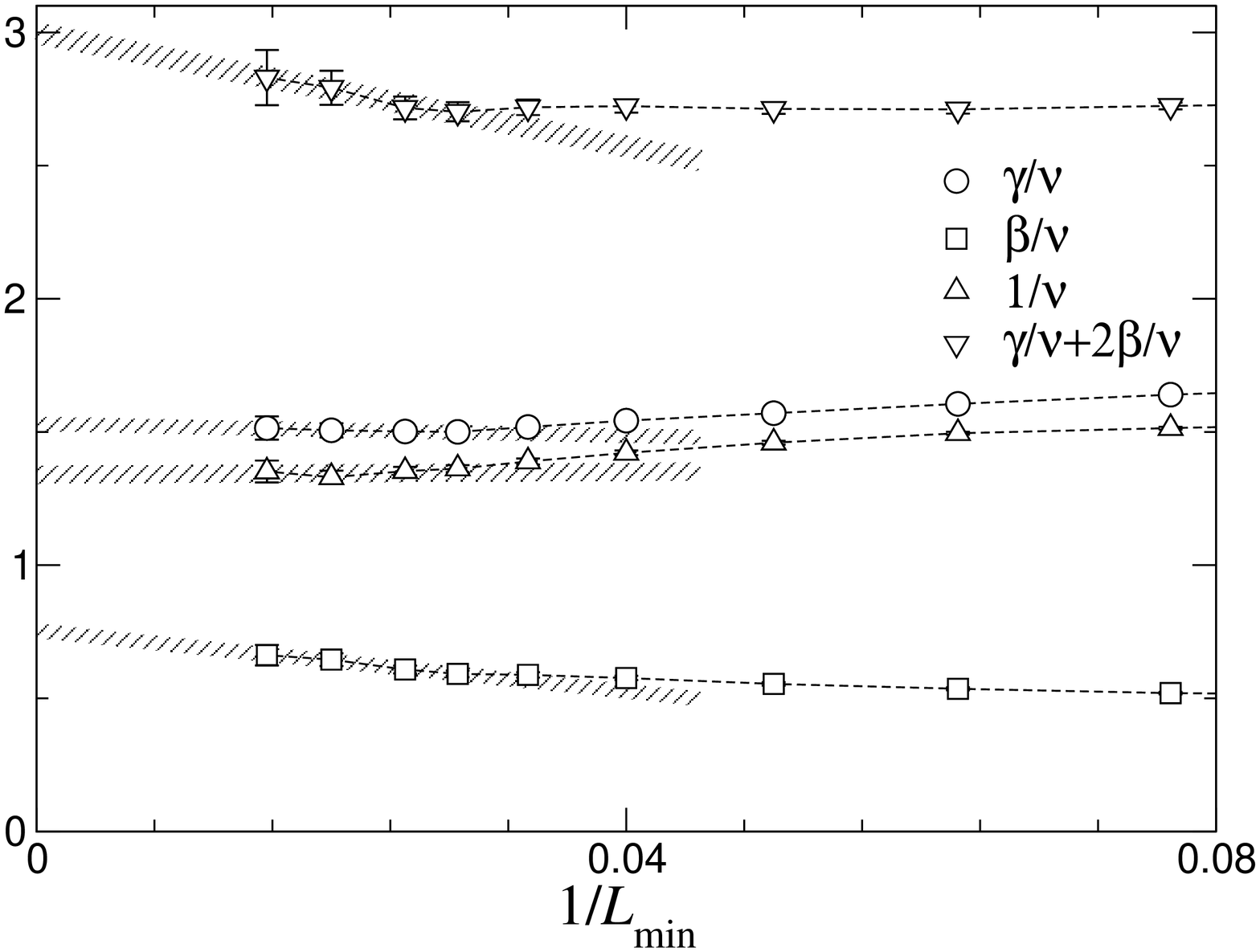}}
        \end{center}\vskip 0cm
        \caption{\small Effective critical 
        exponents $\gamma/\nu$, $\beta/\nu$, and $1/\nu$ for the dilution
        $p=0.56$ obtained from fits between $L_{\rm min}$ and
        $L_{\rm max} = 96$ and extrapolated to $L_{\rm min}\to\infty$. 
        In this limit, the
        scaling relation $\gamma/\nu+2\beta/\nu=D$ is nicely satisfied.}
        \label{FigExpstEffp0.56}
\efig

\subsection{Corrections to scaling}
For the 3D disordered Ising model
it is well known that the correction-to-scaling 
close to the random fixed point 
are strong (with a 
corrections-to-scaling exponent around $\omega=0.4$). 
Let us assume here also the existence
of an irrelevant scaling field $g$ with scaling dimension $y_g=-\omega<0$.
The scaling expression for the susceptibility
\begin{equation}
        \overline\chi(L^{-1},\beta-\beta_c,g)=L^{\gamma/\nu}f_\chi
(L|\beta-\beta_c|^\nu,L^{-\omega}g),
        \label{eq-IrrelevantChi}
\end{equation}
expanded at $\beta_c$ (on a finite system the susceptibility is always finite) 
around the fixed point value $g=0$, leads to the
standard expression 
$\Gamma_c L^{\gamma/\nu}[1+b_\chi L^{-\omega}+O(L^{-2\omega})]$.
In order to investigate this
question for the 3D 4-state Potts model, we tried to fit the
physical quantities for $p=0.56$ as 
\begin{equation}
        \overline\chi_{\rm max}(L)=\Gamma_c L^{\gamma/\nu}(1+b_\chi L^{-\omega}),
        \label{eq-CorrScalChi}
\end{equation}
and similar expressions for $\overline m_{\beta_{\rm max}}$,
in the range $L\ge 25$ where the leading term was already fitted in the
previous section, and the subleading correction is due to the first 
irrelevant scaling field.

Since four-parameter non-linear fits are not stable, we preferred linear
fits where the exponents are taken as fixed parameters but the amplitudes are
free. In Fig.~\ref{FigChi2}, we show a 3D plot of the cumulated 
square deviation of the least-square linear fit, $\chi^2$, as 
a function of $\gamma/\nu$ and $\omega$. There is a clear valley
which confirms that $\gamma/\nu$ is close to 1.5, but the valley is so
\begin{figure}[t]
        \epsfysize=6.3cm
        \begin{center}
        \mbox{\epsfbox{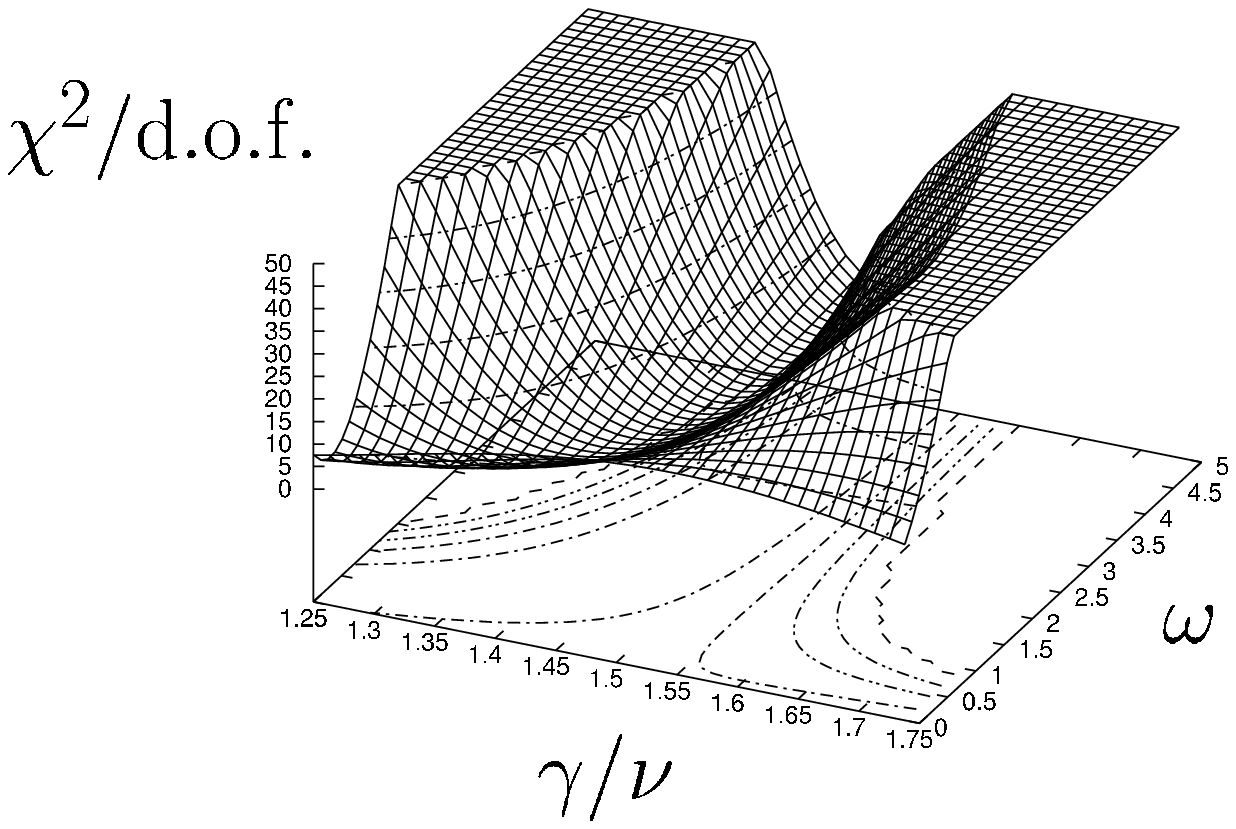}}
        \end{center}
        \caption{\small Plot of the $\chi^2$ deduced from linear fits of
        $\overline\chi_{\rm max}(L)=
        \Gamma_c L^{\gamma/\nu}(1+b_\chi L^{-\omega})$
        in the range $25\leq L\leq 96$ for $p=0.56$.
        The exponents are treated as fixed parameters and the amplitudes are
        free. The base plane gives the ranges of variation of the exponents:
        $1.25\leq\gamma/\nu\leq 1.75$ and $0\leq\omega\leq 5$. The absolute
        minimum is at $\gamma/\nu=1.49$, $\omega=3.88$, but the valley
        is extremely flat in the $\omega$-direction. A cutoff at $\chi^2=50$ 
        has been introduced
        in order to improve clarity of the figure.
        }
        \label{FigChi2}  
\end{figure}
flat in the $\omega$-direction that there is no
clear minimum 
to give a reliable estimation of the
corrections-to-scaling exponent.
The same procedure for $\beta/\nu$  is illustrated
in the next figure (Fig.~\ref{FigChi2M}). Again, there is no way to get a 
compatible
corrections-to-scaling exponent from the three fits, but the leading exponents
are indeed close to $\beta/\nu=0.71$ (and $1/\nu=1.35$).
Of course the minima of $\chi^2$ do not exactly coincide with the data 
presented in the table which should correspond to $\omega\to\infty$.

\begin{figure}
        \epsfysize=6.3cm
        \begin{center}
        \mbox{\epsfbox{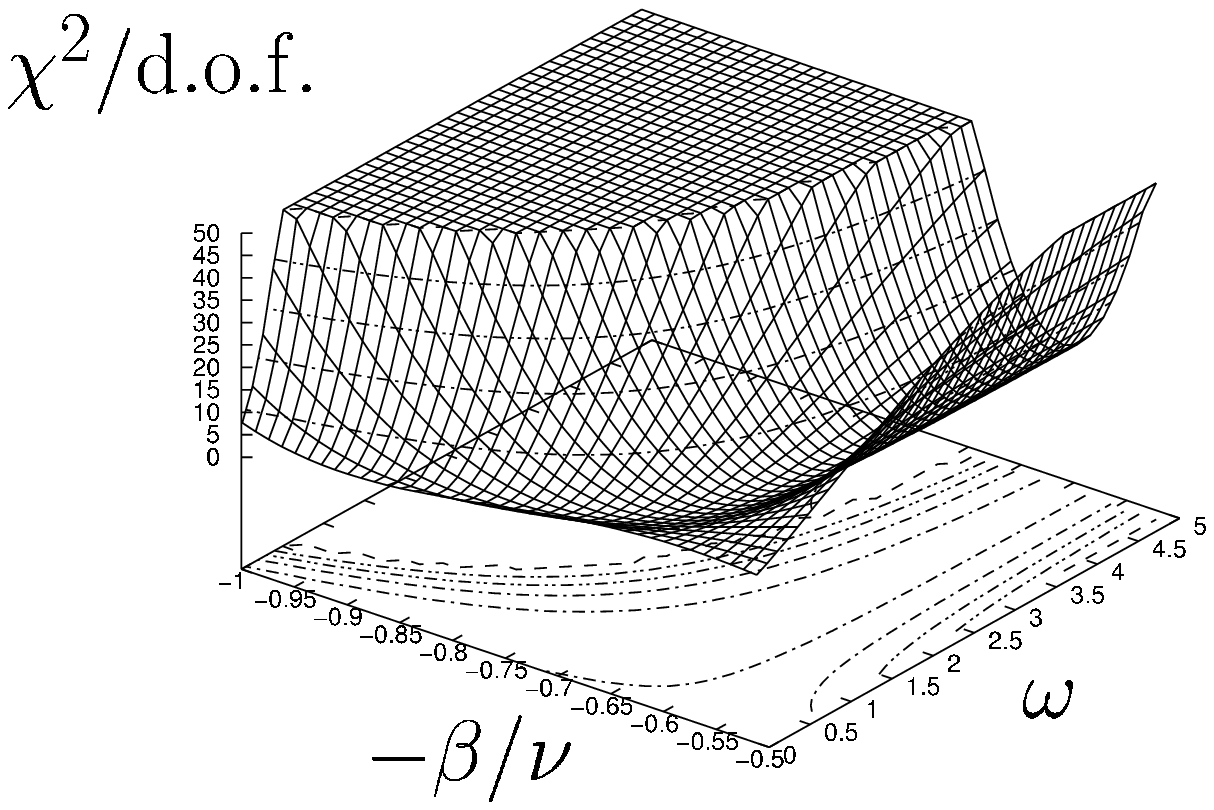}}
        \end{center}
        \caption{\small Plot of the $\chi^2$ deduced from linear fits of
        $\overline m$ (the exponent is thus negative)
        in the range $25\leq L\leq 96$ for $p=0.56$. In the base plane,  
        the range of variation of the exponents is 
        $-1\leq -\beta/\nu\leq -0.5$ and $0\leq\omega\leq 5$, and the 
        minimum is
        at $\beta/\nu=0.85$, $\omega=0.135$.}
        \label{FigChi2M}  
\end{figure}

\section{Conclusion\label{sec7}}
We studied the three-dimensional bond-diluted
4-state Potts model by large-scale
Monte Carlo simulations. The pure system undergoes a strong first-order phase
transition. The numerical estimates of the dynamical exponent $z$ and
of the interface tension give evidences for the existence of a disorder-induced
tricritical point for bond dilutions between $p=0.68$ and $p=0.84$ below which
the transition is softened to second order.
Very strong crossover corrections are observed up to lattice size
$L\le 30-40$. The regime of the
random fixed point is best observed for the bond concentration $p=0.56$. 
From the values of the ratios of exponents measured at that
concentration, 
\begin{eqnarray}
& \gamma/\nu & =1.535(30),\\
& \beta/\nu  & =0.732(24),\\
& 1/\nu      & =1.339(25),
\end{eqnarray}
the
following estimates of the critical exponents are derived: 
\begin{eqnarray}
& \gamma & =1.146(44),\\
& \beta  & =0.547(28),\\
& \nu    & =0.747(14).
\end{eqnarray}
Let us
mention that these exponents are in reasonably good agreement with
recent star-graph high-temperature expansions~\cite{HellmundJanke02} 
of this model which
give $\gamma=1.00(3)$.
The value of $\nu$ is eventually safe with respect to the bound
$\nu\ge 2/D = 0.6666\dots$ of
the stability of the random fixed point.
In the random fixed point regime, we are unable to extract from the numerical
data any reliable correction-to-scaling exponent (linked to the possible 
appearance of irrelevant scaling fields), even though it is clear that such
corrections cannot be ignored.

In some sense, the outcome of this time-consuming work is disappointing, 
since we were not able to reach the asymptotic regime where exponents in
the second-order regime of the phase diagram become dilution-independent,
since the corrections to scaling are too strong, and
since the tricritical point was not located with precision.
We believe that this is due to the extreme difficulty of the problem and
not to an unadapted approach. Perhaps we were too ambitious, but we
have the feeling that the final values given for the critical exponents
are reliable enough and should not be contradicted in the future by
similar studies.

\section*{Acknowledgements}
This work was partially supported by the PROCOPE exchange programme of DAAD 
and EGIDE, the EU-Network HPRN-CT-1999-000161 ``EUROGRID: {\sl Discrete Random
Geometries: From solid state physics to quantum gravity}", the DFG, and the 
German-Israel-Foundation (GIF) under grant No.\ I-653-181.14/1999. We
gratefully acknowledge the computer-time grants 
hlz061 of NIC, J\"ulich, 
h0611 of LRZ, M\"unchen,
062 0011 of CINES, Montpellier
and
2000007 of CRIHAN, Rouen, 
which were essential for this project. The authors gratefully thank Ian Campbell
for a critical reading of the preprint which helped to improve it.

\end{document}